\documentclass[11pt]{article}

\usepackage[margin=1in]{geometry}
\usepackage{graphicx}
\usepackage{amsmath}
\usepackage{natbib}
\usepackage{dsfont}

\usepackage[ruled,linesnumbered]{algorithm2e}
\SetEndCharOfAlgoLine{}

\usepackage{amsthm}
\usepackage{amsfonts}
\usepackage{multirow}
\usepackage{multicol}
\usepackage{pdflscape}
\usepackage{array}

\usepackage{amsmath,bbm,graphicx,amssymb,amsthm,amsbsy}
\usepackage{hyperref}
\hypersetup{
    colorlinks=true,
    linkcolor=blue,
    filecolor=magenta,      
    urlcolor=cyan,
    pdfpagemode=FullScreen,
    }
\usepackage{tabularx,booktabs}
\usepackage[utf8]{inputenc}
\usepackage{url}
\usepackage{enumitem}
\setlist[enumerate]{nosep}
\setlist[itemize]{nosep}

\newcommand{\argmin}{\operatornamewithlimits{argmin}}
\newcommand{\dd}{\text{d}}
\newcommand{\balpha}{{\boldsymbol{\alpha}}}
\newcommand{\btalpha}{\boldsymbol{\tilde\alpha}}
\newcommand{\balphas}{{\boldsymbol{\alpha}}^*}
\newcommand{\balphad}{{\boldsymbol{\alpha}}^\dagger}
\newcommand{\bbeta}{{\boldsymbol{\beta}}}
\newcommand{\btheta}{\boldsymbol{\theta}}
\newcommand{\bb}{\boldsymbol{b}}

\newcommand{\bB}{\boldsymbol{B}}
\newcommand{\bX}{\boldsymbol{X}}
\newcommand{\bY}{\boldsymbol{Y}}
\newcommand{\bSigma}{\boldsymbol{\Sigma}}

\newtheorem{assumption}{Assumption}
\newtheorem{definition}{Definition}
\newtheorem{theorem}{Theorem}
\newtheorem{lemma}{Lemma}
\newtheorem*{remark}{Remark}

\makeatother

\begin{document}

\title{
Semi-Parametric Inference for Doubly Stochastic Spatial Point Processes: An Approximate Penalized Poisson Likelihood Approach

}
\author{Si Cheng\textsuperscript{1}
\and Jon Wakefield\textsuperscript{1,2}
\and Ali Shojaie\textsuperscript{1,2}}
\date{
\textsuperscript{1} Department of Biostatistics, University of Washington, Seattle WA, USA\\
\textsuperscript{2} Department of Statistics, University of Washington, Seattle WA, USA
}

\maketitle
\begin{abstract}

Doubly-stochastic point processes model the occurrence of events over a spatial domain as an inhomogeneous Poisson process conditioned on the realization of a random intensity function. They are flexible tools for capturing spatial heterogeneity and correlation. However, {existing} implementations of doubly-stochastic spatial models are computationally demanding, often have limited theoretical guaranties, and/or rely on restrictive assumptions. 
We propose a penalized regression method for {estimating covariate effects in} doubly-stochastic point processes that is computationally efficient and does not require a parametric form or stationarity of the underlying intensity. 
{Our approach is based on an approximate (discrete and deterministic) formulation of the true (continuous and stochastic) intensity function.
We show that consistency and asymptotic normality of the covariate effect estimates can be achieved despite the model misspecification,}
and develop a covariance estimator that leads to {valid, albeit} conservative, statistical inference. 
A simulation study shows the validity 
of our approach under less restrictive assumptions on the data generating mechanism. An application to Seattle crime data demonstrates better prediction accuracy and narrower confidence intervals 
compared to existing alternatives.

\emph{Keywords}: Cox process, spatial point process, semi-parametric model, high-dimensional inference, non-stationarity
\end{abstract}

\section{Introduction}
\label{sec:intro}

Spatial point process models \citep{diggle2003statistical,moller2003statistical,illian2008statistical,chiu2013stochastic} are widely used to study spatial event patterns in epidemiology \citep{best2005comparison,franch2020spatial}, sociology \citep{ferreira2012gis,leong2015review}, and ecology \citep{law2009ecological,renner2015point}.
Two structural features frequently arise in these applications: \emph{spatial heterogeneity} and \emph{spatial correlation} \citep{anselin1988spatial,plotkin2000species,vinatier2011factors}.
Spatial heterogeneity refers to systematic variation in the intensity of events across space,
often attributable to covariate effects or large-scale baseline variation,
whereas spatial correlation captures residual clustering or dependence beyond first-order effects.

Doubly stochastic Poisson processes, also known as Cox processes \citep{cox1955some}, provide a natural mechanism for modeling both phenomena by introducing a latent random intensity field.
Conditional on this latent field, the outcome events follow a Poisson process, and spatial dependence is induced through stochastic variation in intensity \citep{moller2007modern}.
A canonical example is the log-Gaussian Cox process (LGCP), in which the log-intensity is modeled as a Gaussian random field \citep{moller1998log,diggle2013spatial}.
While Cox processes offer substantial modeling flexibility, their likelihood-based inference is analytically intractable due to the infinite-dimensional latent intensity.
Consequently, practical inference relies on discretization, basis expansion, or other functional approximations.

Grid-based MCMC, Hamiltonian Monte Carlo, and related simulation-based methods remain standard tools for LGCP inference \citep{moller1998log, moller2003statistical, teng2017bayesian}.
Deterministic approximations, such as integrated nested Laplace approximation (INLA) and stochastic partial differential equation (SPDE) representations, provide a scalable alternative to simulation-based posterior computation \citep{rue2009approximate,lindgren2011explicit,simpson2016going}.
Discretization for likelihood approximation has been particularly common in aggregated count settings \citep{li2012log, johnson2019spatially}, and more recent work combines reduced-rank representations with variational approximations to optimize marginal likelihoods \citep{dovers2023fast}.
These methods are powerful for recovering latent intensity surfaces and covariance parameters, and for uncertainty quantification under parametric latent-field assumptions. However, inference for regression effects remains inherently model-based: validity depends on correct specification (or sufficiently accurate approximation) of the latent stochastic structure. 

A complementary literature 
avoids direct evaluation of the likelihood by using estimating functions 
derived from moment identities. A key insight is that the Poisson likelihood score yields an unbiased estimating function for regression parameters under broad classes of spatial point processes, even when the true mechanism exhibits spatial clustering \citep{schoenberg2005consistent, guan2007thinned}. 
Composite likelihood approaches construct tractable surrogate objectives from lower-order components and have been used to fit spatial point process models without full likelihood evaluation \citep{guan2006composite,baddeley2017local}. Quasi-likelihood and weighted estimating equation approaches further incorporate known second-order dependence properties to improve efficiency 
\citep{guan2010weighted, diggle2010estimating,deng2017second,chu2022quasi}. 
{The asymptotic properties of these methods are more tractable, even under spatial correlation; however, inference and efficiency improvements may depend sensitively on second-order assumptions or stationarity conditions.} 

Several extensions, most notably multi-stage composite likelihood procedures, to nonstationary and inhomogeneous settings have been proposed \citep{waagepetersen2007estimating,waagepetersen2009two,dvovrak2019quick}. While these approaches broaden applicability, statistical inference is not always explicit, or remains tied to second-order assumptions and/or relatively low-dimensional parameter regimes.


In parallel, a growing literature develops regularization methods for spatial point process regression, addressing modern settings with high-dimensional covariates.
Regularized estimating equations based on Poisson and logistic-regression-type criteria have been proposed for feature selection and intensity estimation, with consistency guarantees under suitable conditions \citep{choiruddin2018convex,choiruddin2023adaptive,rakshit2021variable}.
While these methods provide an important theoretical foundation for high-dimensional estimation in point process models, comparatively less is known about post-regularization inference for regression effects under doubly stochastic mechanisms, particularly when discretization is used and the working Poisson model is misspecified due to latent-field-induced overdispersion.


Motivated by the ubiquity of discretization in Cox process estimation and the limitations of existing inference procedures, 
especially with high-dimensional covariates, we propose a semi-parametric framework that approximates a Cox process through a penalized Poisson model fitted to discretized counts. Our primary objective is consistent estimation and valid inference for covariate effects under doubly stochastic spatial models, a central task in epidemiology \citep{mahaki2011multivariate,li2012log}, environmental health \citep{jerrett2005spatial}, and sociology \citep{rostami2017modeling,adeyemi2021demography}.

{To summarize, existing methods leave open the problem of conducting scalable high-dimensional inference in doubly stochastic spatial models, without specifying the latent field or its second-order dependence structure.} Our contributions bridge this gap and  differ from existing penalized likelihood or estimating-equation approaches in three key respects:
\begin{itemize}
    \item \emph{Working-model robustness.} We treat the discretized Poisson likelihood as a working model and establish consistency and asymptotic normality for regression parameters when the data arise from a Cox process and the latent intensity is ignored. We do not require a parametric pair correlation function or stationarity of the latent field.
    \item \emph{High-dimensional inference via debiasing.} We combine penalized Poisson maximum likelihood (PMLE) estimation with de-biasing to enable valid inference when the number of covariates grows with the observation domain, extending high-dimensional generalized linear model (GLM) inference to spatially dependent and misspecified settings.
    \item \emph{Computational scalability.} Inference reduces to penalized Poisson regression on discretized cells, with scalability comparable to standard penalized GLM solvers.
\end{itemize}

{Section~\ref{sec:setting} introduces the model setting. To develop our proposed approach,} in Section~\ref{sec:model} we approximate the analytically intractable Cox process by discretizing the observation window and modeling counts within cells via a Poisson regression framework. This yields a tractable PMLE problem analogous to \citet{schoenberg2005consistent}, but adapted to high-dimensional settings and to doubly stochastic mechanisms.
{We justify the specification of a working Poisson model by establishing the consistency of the regression parameter estimates} under an asymptotic regime similar to increasing domain asymptotics.
{Under a few additional assumptions,} in Section~\ref{sec:inference} we establish the asymptotic normality of de-biased estimates of these parameters {accounting for the randomness ignored in first-order modeling}. 
The performance of our approach is illustrated and compared with state-of-the-art methods via simulation studies in Section~\ref{sec:sim}, as well as an application to Seattle crime data in Section~\ref{sec:data}. 
\section{Problem Setting}
\label{sec:setting}

Consider a Cox process $\mathcal Y(s):s\in\Omega$ over an observation window
$\Omega$. Conditional on a realized intensity function $\lambda(\cdot)$,
$\mathcal Y(\cdot)$ is an inhomogeneous Poisson process, where
$\lambda(\cdot)$ is a realization of the random intensity $\Lambda(\cdot)$
modeled as
\begin{equation}
    \log \Lambda(s)
    =
    \log P(s) + \alpha^0(s) + X(s)\boldsymbol\beta^0 + \varepsilon(s).
    \label{equ:intensity}
\end{equation}
Here, $P(s)$ is an offset, $\alpha^0(s)$ is the baseline log-intensity,
$X(s)$ is a $p$-dimensional covariate function, $\boldsymbol\beta^0\in\mathbb R^p$
is the slope parameter of interest, and $\varepsilon(s)$ is a mean-zero latent error
field. For example, a Gaussian random field $\varepsilon(s)$ yields a LGCP.


\citet{schoenberg2005consistent} shows that maximizing a Poisson likelihood
can yield consistent estimates for certain low-dimensional, parametric,
non-Poisson point processes. Extending this idea to Cox processes with
high-dimensional covariates can greatly simplify the otherwise less tractable
optimization problem. Following \citet{schoenberg2005consistent}, we refer to
$\lambda(\cdot)$ as the \textit{conditional intensity} and to the pointwise expectation
$\mathbb E_0[\Lambda(\cdot)]$ under the data-generating mechanism with respect to $\varepsilon(\cdot)$ as the \textit{unconditional intensity}. The latter satisfies
\begin{equation}
\mathbb E_0[{\Lambda(s)}]
=
P(s)\exp\left[\alpha^0(s)+X(s) \boldsymbol\beta^0+\phi(s)\right],
\ \text{where }
\phi(s)=\log\mathbb E_0\left[\exp\{\varepsilon(s)\}\right].
\label{equ:uncond-intensity}
\end{equation}

Although the point process is defined on the continuous domain $\Omega$, events
and covariates are often observed only through region-level counts or summary
statistics. Let $\Omega=\Omega_1\cup\cdots\cup\Omega_n$ denote the observed
spatial partition. The aggregated counts then satisfy
\begin{align}
    Y_i\mid \lambda_i
    &\sim
    \operatorname{Poisson}(\lambda_i),
    \qquad i=1,\ldots,n,
    \notag\\
    \lambda_i\mid \varepsilon(\cdot)
    &=
    \int_{\Omega_i}
    P(s)
    \exp\left[
        \alpha^0(s)
        +
        X(s)\boldsymbol\beta^0
        +
        \varepsilon(s)
    \right]
    \dd s.
    \label{equ:model}
\end{align}
Here $Y_i$ is the event count in $\Omega_i$, while
$\alpha^0(\cdot)$, $P(\cdot)$, $X(\cdot)$, and $\varepsilon(\cdot)$ are the
continuous fields in \eqref{equ:intensity}. In practice, model fitting usually
uses a region-level offset $P_i$ and covariate vector $X_i\in\mathbb R^p$,
as is common when spatial outcomes are aggregated for confidentiality or
combined with non-spatial information \citep[see, e.g., Section~\ref{sec:data},
and][for further examples and discussion]{diggle2010estimating,li2012log,taylor2018continuous}.


{Taking expectation with respect to the latent field, and interchanging expectation and integration under the conditions discussed in Section~\ref{subsec:consistency},
we obtain
\begin{align}
    \mathbb E_0[Y_i\mid X(\cdot)]
    &=
    \mathbb E_0[\lambda_i\mid X(s)] \notag =
    \int_{\Omega_i} P(s)
    \exp[\alpha^0(s) + X(s)\bbeta^0]
    \mathbb E_0[\exp\{\varepsilon(s)\}]
    \dd s \notag\\
    &=
    \exp\left(X_i\boldsymbol\beta^0\right)
    \int_{\Omega_i} P(s) 
    \exp\left[\alpha^0(s)+ \left(X(s) - X_i\right)\bbeta^0 + \phi(s)\right]
    \dd s.
    \label{equ:uncond-cell-mean}
\end{align}
}

{In Section~\ref{subsec:consistency} we shall see that the unconditional intensity \eqref{equ:uncond-intensity} and discretized cell means \eqref{equ:uncond-cell-mean} are key quantities connecting the working Poisson log-likelihood and parameters underlying the more complex Cox process model. Although the fitted model uses region-level $P_i$ and
$X_i$, our theory allows the event process, covariates, baseline log-intensity,
and latent error field to vary continuously over $\Omega$. The resulting
discretization error, including the $(X(s)-X_i)\bbeta^0$ term in
\eqref{equ:uncond-cell-mean}, is controlled under the regularity conditions in
Section~\ref{subsec:consistency}.


\section{Penalized Poisson Maximum Likelihood Estimation 
}
\label{sec:model}

\subsection{Model Specification}
We aim to estimate and conduct inference on
${\boldsymbol\beta^0}\in \mathbb R^p$ from the observed counts
$\boldsymbol{Y}:=(Y_1,\ldots,Y_n)^\top$ and region-level covariates
$\boldsymbol{X}:=[X_1^\top,\ldots,X_n^\top]^\top\in\mathbb R^{n\times p}$,
under minimal assumptions on the latent random field $\varepsilon(\cdot)$. Without loss of generality, we take
the known offset $P(s)\equiv 1$; otherwise, it can be absorbed into
$\alpha^0(s)$ in the theoretical formulation.


{We note from (\ref{equ:model}) that the expected case counts $Y_i$ conditioning on $\varepsilon(s)$ is
\begin{align*}
\mathbb E[Y_i\mid\varepsilon(\cdot)] &= |\Omega_i|\exp\left[\tilde\alpha_i^0(\varepsilon(\cdot)) + X_i\bbeta^0\right], 
\text{ where}\\ 
\tilde\alpha_i^0(\varepsilon(\cdot)) &:= \log\left[{|\Omega_i|^{-1}}\int_{\Omega_i}\exp\left[\alpha^0(s) + {\left(X(s)-X_i\right)\bbeta^0} + \varepsilon(s)\right] \ \dd s\right],
\end{align*}
{and $|\Omega_i|$ is the area of $\Omega_i$.} 
This resembles a Poisson mixed
effect model, except that the spatial random effects
$\tilde\alpha_i^0$ have unknown and generally intractable distributions.
Motivated by this connection, we specify a working Poisson regression model
with discretized region-specific baselines
$\tilde{\boldsymbol\alpha}\in\mathbb R^n$, where the tilde distinguishes this
discretized vector from the continuous baseline intensity function
$\alpha^0(s)$. Using the same discretization for $\boldsymbol X$ and
$\boldsymbol Y$, the working Poisson log-likelihood is
\begin{equation}
    \ell(\tilde{\boldsymbol{\alpha}}, {\boldsymbol\beta};\boldsymbol X, \boldsymbol Y)
    =
    \sum_{i=1}^n y_i\left(\tilde\alpha_i + X_i \boldsymbol\beta\right)
    -
    \sum_{i=1}^n |\Omega_i|
    \exp\left(\tilde\alpha_i + X_i\boldsymbol\beta\right).
    \label{equ:pois-loglik}
\end{equation}
Unlike Poisson mixed models, we do not impose a parametric distribution on
$\tilde\alpha_i$, but model their realization as fixed region-specific
parameters, allowing the working model to apply to a broader class of point
processes.
The working likelihood \eqref{equ:pois-loglik} involves two approximations to the true mechanism: discretizing the continuous fields $\alpha^0(s)$, $X(s)$ and
$\varepsilon(s)$, and absorbing the latent-field randomness into
$\tilde{\boldsymbol\alpha}$. This approximation is motivated by
\citet{schoenberg2005consistent}, which we extend to high-dimensional and
semiparametric Cox processes in Section~\ref{subsec:consistency}. In
particular, we show that the gradient of \eqref{equ:pois-loglik} yields a valid
estimating equation for $\boldsymbol\beta$ despite the misspecified discrete
form of $\tilde{\boldsymbol\alpha}$ and the ignored randomness from
$\varepsilon(\cdot)$.


Because both $\tilde{\boldsymbol\alpha}$ and $\boldsymbol\beta$ are
high-dimensional, we impose penalties to ensure identifiability for this over-parameterized model. This is
related to penalized regression for mixed effects models
\citep[see, e.g.,][]{heckman2013penalized}, although our penalties need not match the latent covariance structure. We use an $\ell_1$ sparsity penalty on
$\boldsymbol\beta$ \citep{tibshirani1996regression}, and an additional $\ell_1$
\citep{tibshirani2005sparsity} or $\ell_2$
\citep{zhao2016grace, li2019prediction} fusion penalty on
$\tilde{\boldsymbol\alpha}$.

More specifically, the partition of $\Omega$ induces a graph
$\mathcal G_n=(V_n,E_n)$, with vertices
$V_n=\{\Omega_1,\ldots,\Omega_n\}$ and edges $E_n \subseteq V_n\times V_n$ connecting adjacent regions.
The edges may be unweighted or weighted by centroid distances or other measures of
(dis)similarity. Let $W_n=(w_{ij})$ be the adjacency matrix and
$D_n=\operatorname{diag}(d_1,\ldots,d_n)$, where
$d_i=\sum_{j\in V_n}w_{ij}$. The edge incidence matrix
$B_n\in\mathbb R^{|E_n|\times |V_n|}$ has, for an edge
$(\Omega_i,\Omega_j)$ with $i<j$, entries
$b_{ki}=\sqrt{w_{ij}}$ and $b_{kj}=-\sqrt{w_{ij}}$ in its $k$th row. The graph
Laplacian \citep{chung1997spectral} is
$L_n=D_n-W_n=B_n^\top B_n$. Since $L_n\mathbf 1=0$ where $\mathbf{1}$ is a vector of all ones, we follow
\citet{li2019prediction} and use
$\tilde L_n:=L_n+\delta I_n$ with a small $\delta>0$ to enforce numerical stability.
The fusion penalty is

\begin{equation*}
    R(\boldsymbol{\tilde\alpha};\mathcal G_n) = 
    \begin{cases}
        \lVert B_n\boldsymbol{\tilde\alpha}\rVert_1
    = \sum_{(\Omega_i,\Omega_j)\in E_n} \sqrt{w_{ij}}\lvert \tilde\alpha_i - \tilde\alpha_j \rvert \hspace{27.5mm} (\ell_1)\\
        \frac{1}{2} \boldsymbol{\tilde\alpha}^\top \tilde L_n \boldsymbol{\tilde\alpha}
    = \frac{1}{2}\sum_{(\Omega_i,\Omega_j)\in E_n} w_{ij}(\tilde\alpha_i - \tilde\alpha_j)^2 + \frac{\delta}{2} \sum_{i=1}^n\tilde\alpha_i^2 \quad (\ell_2)
    \end{cases}.
\end{equation*}

The $\ell_1$ fusion penalty, which is a generalized Lasso penalty
\citep{tibshirani2011solution}, encourages a piecewise constant baseline over
connected regions, whereas the $\ell_2$ penalty encourages smooth but not
exactly equal neighboring baselines.

The penalized PMLE is given by the solution to the optimization problem 
\begin{equation}
    \boldsymbol{\hat\theta}:=\left(\hat{\boldsymbol\alpha}^\top, \hat{\boldsymbol\beta}^\top\right)^ \top =
    \argmin_{\tilde{\boldsymbol\alpha},\boldsymbol\beta}
    -\ell(\tilde{\boldsymbol{\alpha}}, {\boldsymbol\beta};\boldsymbol X, \boldsymbol Y)
    + \gamma_n R(\tilde{\boldsymbol{\alpha}};\mathcal G_n) +\tau_n\lVert\boldsymbol\beta\rVert_1,
    \label{equ:obj-fun}
\end{equation}
where $\gamma_n$ and $\tau_n$ are tuning parameters, for example selected by
cross-validation. Prediction and cross-validation strategies under regional dependence are
discussed in Section~\ref{subsec:pred}.

The proposed penalized PMLE is also related to Bayesian spatial models with
intrinsic conditional autoregressive priors \citep{besag1974spatial}. Under
the $\ell_2$ fusion penalty, it resembles the maximum a posteriori estimate of the
Besag-York-Molli{\'e} model \citep{besag1991bayesian}, whose spatially
correlated random effects follow a Gaussian Markov random field. 
Its quadratic prior form parallels the $\ell_2$ fusion penalty in
\eqref{equ:obj-fun}, although our formulation treats
$\tilde{\boldsymbol\alpha}$ as fixed nuisance parameters rather than requiring
a full parametric random-effects distribution.


\subsection{{Optimization Algorithm}}
\label{subsec:comp}

{The optimization procedure to solve \eqref{equ:obj-fun} differs from standard convex optimization algorithms due to the inclusion of fusion and sparsity penalties.}
Defining the soft-thresholding operator $S_\tau(x):= \text{sign}(x)\max\{|x| - \tau, 0\}$ {for $0<\tau<\infty$}, the optimization {involving only the sparsity penalty $\tau_n\lVert\bbeta\rVert_1$} can be {solved} using the proximal gradient descent algorithm {\citep{beck2009fast, parikh2014proximal}}.
{The convex and smooth $\ell_2$ fusion penalty would not introduce additional complexity to the optimization. However,} with the $\ell_1$ fusion penalty, $R(\boldsymbol{\tilde\alpha};\mathcal G_n)$ is nonseparable with respect to $\boldsymbol{\tilde\alpha}$, which introduces challenges in optimization for nonlinear models such as the Poisson model. To overcome these challenges, we follow \citet{chen2012smoothing} and adopt a smooth $\ell_\infty$ approximation for the $\ell_1$ fusion penalty,
\begin{equation}
    \gamma_n\lVert B_n \boldsymbol{\tilde\alpha}\rVert_1 \approx h_{\xi}(\boldsymbol{\tilde\alpha}):=
    \gamma_n \max_{\lVert\nu\rVert_\infty \le 1}
    \left[ \nu^\top B_n\boldsymbol{\tilde\alpha} - \frac{\xi}{2}\lVert\nu\rVert_2^2 \right].
    \label{equ:smooth-l1}
\end{equation}
The parameter $\xi$ controls the amount of smooth relaxation, with $\xi = 0$ recovering the original $\ell_1$ fusion penalty. The gradient of $h_\xi$ can simply be calculated as 
\[
    \nabla h_\xi(\boldsymbol{\tilde\alpha}) = \gamma_n  B_n^\top S_\infty\left(\frac{B_n\boldsymbol{\tilde\alpha}}{\xi}\right),
\]
where $S_\infty(\cdot)$ is the element-wise projection operator onto the $\ell_\infty$ ball, {i.e., $S_\infty(x) = x$ for $|x| \le 1$ and $S_\infty(x) = \mathrm{sign}(x)$ for $|x| > 1$}.
{Incorporating the proximal gradient descent along with the smooth approximation \eqref{equ:smooth-l1}, we define our optimization procedure in Algorithm~\ref{algo:l2}.}
\citet{chen2012smoothing} show that with $\xi = \epsilon/|E_n|$, the approximation gap $\left\lvert \gamma_n\lVert B_n\boldsymbol{\tilde\alpha}\rVert_1 - h_{\xi}(\boldsymbol{\tilde\alpha}) \right\lvert\le\epsilon$ is guaranteed within $O\big(\sqrt{|E_n|}/\epsilon\big)$ iterations.
{In} Algorithm~\ref{algo:l2}, the step size is set adaptively via backtracking line search {\citep{armijo1966minimization, boyd2004convex}}.
Lines 2 through 11 can be replaced by coordinate-wise gradient descent which may facilitate the tuning of $\gamma_n$, $\tau_n$. 

\RestyleAlgo{boxruled}
\begin{algorithm}[ht]
  \caption{Proximal gradient descent for penalized PMLE\label{algo:l2}}
  Set tolerance $tol$ as well as (small) positive constants $a, b$ for backtracking line search. 
  {Set the tuning parameters to be $\gamma_n,\tau_n$}.
  Initialize $\boldsymbol\theta^{(0)}=(\boldsymbol{\tilde\alpha}^{(0)}, \boldsymbol\beta^{(0)})$ and calculate the objective function 
        \[f(\boldsymbol{\theta}^{(0)})= \mathcal L(\boldsymbol{\theta}^{(0)}) + \tau_n\lVert\boldsymbol\beta^{(0)}\rVert_1
        :=
        -\ell(\boldsymbol{\theta}^{(0)}) + \gamma_n R(\boldsymbol{\tilde\alpha}^{(0)}) + \tau_n \lVert\boldsymbol\beta^{(0)}\rVert_1\]\;
        
    \For{$t = 0,1, \ldots$ until convergence}{
        Evaluate the gradient $\nabla\mathcal L(\boldsymbol\theta^{(t)}):= -\nabla\ell(\boldsymbol\theta^{(t)})+ {\nabla \gamma_n R(\boldsymbol{\tilde\alpha};\mathcal G_n)}$, where\;

        {
        \begin{equation*}
        \nabla \gamma_n R(\boldsymbol{\tilde\alpha};\mathcal G_n) \doteq  
        \begin{cases}
            \nabla h_\xi(\boldsymbol{\tilde\alpha}) = \gamma_n B_n^\top S_\infty\left(\frac{ B_n\boldsymbol{\tilde\alpha}}{\xi}\right), \text{ for } \ell_1 \text{ fusion penalty;}\\
            \gamma_n \tilde L_n\boldsymbol{\tilde{\alpha}}^{(t)}, \text{ for } \ell_2 \text{ fusion penalty.}
        \end{cases}.
        \end{equation*}
        }
        
        Line search: 
            set the initial step size $\eta^{(t)}:= 1$\;
            \While{$\mathcal L\left(\boldsymbol\theta^{(t)} -\eta^{(t)}\nabla\mathcal L(\boldsymbol\theta^{(t)})\right) - \mathcal L(\boldsymbol\theta^{(t)}) \ge 
            {-a\eta^{(t)}\lVert\nabla\mathcal L(\boldsymbol\theta^{(t)})\rVert_2^2}$\;
           }{
                $\eta^{(t)} \leftarrow b\eta^{(t)}$\;
            }
            
        Gradient step: 
            $\boldsymbol{\theta}^\dagger:= (\boldsymbol{\tilde\alpha}^\dagger, \boldsymbol{\beta}^\dagger)\leftarrow \boldsymbol\theta^{(t)} - \eta^{(t)}\nabla\mathcal L(\boldsymbol\theta^{(t)})$ \;
        
         Proximal step: 
            $\boldsymbol\theta^{(t+1)}\leftarrow \left( \boldsymbol{\tilde\alpha}^\dagger, S_{\tau_n}(\boldsymbol{\beta}^\dagger) \right)$ where $S_{\tau_n}(\cdot)$ is applied element-wise on $\boldsymbol{\beta}^\dagger$\;
        
        Convergence criterion:
            Calculate $f(\boldsymbol\theta^{(t+1)})$ and convergence is achieved if $\left\lvert f(\boldsymbol\theta^{(t+1)}) - f(\boldsymbol\theta^{(t)}) \right\rvert < tol\cdot\left\lvert f(\boldsymbol\theta^{(t)}) \right\rvert $
    }
    \KwResult{Output $\boldsymbol\theta^{(t+1)}$}
\end{algorithm}

\subsection{Prediction {for Unobserved Regions}}
\label{subsec:pred}
Although our primary focus is estimation and inference for high-dimensional covariate effects $\bbeta$, out-of-sample prediction of aggregated event counts over small regions is useful for cross-validation and for predicting in new or missing-data regions. Since $\alpha(\cdot)$ is approximated by discretized region-specific baselines $\boldsymbol{\tilde\alpha}$, this requires predicting baselines for test regions. We use the $\ell_2$ cohesion approach of \citet{li2019prediction}. Let $n_1$ and $n_2$ denote the numbers of training and test regions, and partition the graph Laplacian over all $n=n_1+n_2$ regions as
\[
    L_n = \begin{bmatrix}
        L_{11} & L_{12}\\
        L_{21} & L_{22}
    \end{bmatrix},
\]
where $L_{11}$ and $L_{22}$ correspond to the training and test regions. Partitioning
$\boldsymbol{\tilde\alpha}=(\boldsymbol{\tilde\alpha}_1^\top,\boldsymbol{\tilde\alpha}_2^\top)^\top$ and setting $\boldsymbol{\tilde\alpha}_1:=\boldsymbol{\hat\alpha}_1$ obtained from model-fitting, the test-region baselines are predicted by
\[
    \boldsymbol{\hat\alpha}_2 = \argmin_{\boldsymbol{\alpha}} 
    \left(\boldsymbol{\hat\alpha}_1^\top, \boldsymbol{\alpha}^\top\right)
    L_n 
    \left(\boldsymbol{\hat\alpha}_1^\top, \boldsymbol{\alpha}^\top\right)^\top
    = -L_{22}^{-1}L_{21}\boldsymbol{\hat\alpha}_1,
\]
with $\boldsymbol{\hat\alpha}_2=0$ for test regions disconnected from the training sub-graph. Although graph-based train--test splits of dependent samples are not theoretically justified, \citet{li2019prediction} found this procedure to perform reasonably well for cross-validation in practice.

\subsection{{Consistency of Penalized PMLE}}
\label{subsec:consistency}

In this section, we establish consistency of the penalized PMLE in
(\ref{equ:obj-fun}) under $\ell_1$ sparsity and either $\ell_1$ or $\ell_2$
fusion penalties. Throughout, we use the conventional empirical-process
notation
\(\mathbb P_0 f := \mathbb E_0[f]\)
\citep[e.g.,][]{vandervaart1996weakconvergence,vandervaart1998asymptotic},
where \(f\) is a measurable data-dependent function indexed by parameters,
and the expectation is taken under the true data-generating mechanism. Hence
\(-\mathbb P_0\ell(\boldsymbol{\tilde\alpha},\boldsymbol\beta)\)
is the population, or expected, negative working log-likelihood. We first
characterize the target parameter, defined as the minimizer of this population
criterion, and relate it to the true slope parameter $\bbeta^0$ and the
underlying Cox process intensity $\alpha^0(\cdot)$. In particular, we show
that the Poisson likelihood yields an unbiased estimating equation for
$\boldsymbol\beta^0$ despite the ignored random field and the misspecification
of $\alpha^0(\cdot)$. We then use empirical process arguments to prove
convergence of the penalized PMLE to the target parameters under regularization.

Consistency for spatial processes depends on the asymptotic regime. Unlike
independent sampling, spatial data admit multiple regimes under which the same
estimator can behave differently \citep{stein1999interpolation, zhang2005towards}.
We therefore define our regime below, which is related to classical increasing
domain asymptotics.

\begin{definition}[Asymptotic regime]
    Let the observation window $\Omega$, {graph $\mathcal G_n$, covariate surface $X(\cdot)$, baseline intensity surface $\alpha^0(\cdot)$, discretized covariate matrix $\bX$ and outcome $\bY$, the covariate dimensionality $p$, and model parameters $\tilde{\balpha}, \bbeta$ all be explicitly or implicitly indexed by $n$ and vary with $n$. We suppress the subscript $n$ with no confusion arises.}
    
    Let the size of the spatial domain $|\Omega|\rightarrow\infty$ as $n\rightarrow\infty$. The partition $\Omega = \Omega_1\cup\cdots\cup\Omega_n$ satisfies 
    $0 < a_0 \le \liminf_{n\rightarrow\infty}\min_{i = 1, \ldots, n}|\Omega_i| \le \limsup_{n\rightarrow\infty}\max_{i = 1, \ldots, n}|\Omega_i| \le A_0 < \infty $ and the offset satisfies 
    \(
    0<a_0
    \le
    \liminf_{n\rightarrow\infty}
    \min_{i=1,\ldots,n}
    |\Omega_{i}|
    \le
    \limsup_{n\rightarrow\infty}
    \max_{i=1,\ldots,n}
    |\Omega_{i}|
    \le
    A_0
    <\infty,
\)
where $a_0$ and $A_0$ are constants not depending on $n$.

For each region $\Omega_{i}$, let $X_{i}\in\mathbb R^{p}$ denote the
observed region-level covariate vector used in the working likelihood. 
We allow both the covariate dimension $p$ and the sparsity level {$q:=\|\boldsymbol\bbeta^0\|_0$} to grow with $n$, subject to the rate conditions stated in Assumption~\ref{assump:sparse}.
\label{def:asym}
\end{definition}

In words, the observation window expands and incorporates new, unobserved regions as $n$ grows. Correspondingly, the partition includes more and more regions, while maintaining a constant rate of granularity. {This requirement is not restrictive given that we allow $\alpha^0(\cdot)$, {$X(\cdot)$} and $\varepsilon(\cdot)$ to be non-constant within each cell.}
Note that the domain of $\alpha^0(\cdot)$, {$X(\cdot)$} and $\varepsilon(\cdot)$, the discretized covariates $\boldsymbol X$, and the graph $\mathcal G_n=(V_n, E_n)$ induced by the partition all depend on $\Omega$ and $n$. Requirements on their behavior as $n$ increases are stated under our full set of assumptions for consistency, which we now present along with some interpretations.

\begin{assumption}[Regularity conditions]\hfill
    \begin{enumerate}
        \item[\romannumeral1)] The partition $\Omega = \Omega_1\cup\cdots\cup\Omega_n$ is such that each $\Omega_i$ is bounded and connected, and the true baseline function $\alpha^0(\cdot)$ {as well as the covariate surface $X(s)$ are} continuous on each $\Omega_i$.
        \item[\romannumeral2)] The function $\phi(s) := \log\mathbb E_0 \left[\exp\varepsilon(s)\right]$ as defined in (\ref{equ:uncond-intensity}) is continuous on each $\Omega_i$.
        \item[\romannumeral3)] Let $\mathcal F$ be a $\sigma$-algebra over $\Omega$, $\mu(\cdot)$ be a measure (e.g. the Lebesgue measure) defined on $(\Omega,\mathcal F)$ and $\mathbb P_{\varepsilon}$ be the probability measure of the random field $\varepsilon(\cdot)$ defined on $(\Omega_\varepsilon, \mathcal F_\varepsilon)$. Then, there exists a product measure $\rho(\cdot)$ on $(\Omega\times \Omega_\varepsilon, \mathcal F\times\mathcal F_\varepsilon)$ such that for every $A\in\mathcal F$ and $A_\varepsilon\in\mathcal F_\varepsilon$, $\rho(A\times A_\varepsilon) = \mu(A)\mathbb P_\varepsilon(A_\varepsilon)$. We assume that 
        \[
            \lim\sup_{n\rightarrow\infty} \max_{i = 1, \ldots, n} \int_{\Omega_i\times \Omega_\varepsilon} 
            \exp\left[\alpha^0(s) + {X(s)}\boldsymbol\beta^0 + \varepsilon(s) \right] \dd \rho(s,\varepsilon) < \infty.
        \]
    \end{enumerate}
    \label{assump:regularity}
\end{assumption}

Condition \romannumeral3) of Assumption~\ref{assump:regularity} enables the application of Fubini's Theorem, 
so that we only need to learn about functionals of the error random field evaluated pointwise, without explicitly handling the integral involving 
$\varepsilon(\cdot)$.
The combined conditions
further guarantee the existence of {a finite number of locations} 
at which the local unconditional intensity given by (\ref{equ:uncond-intensity}) are representative of the regional mean. This ensures the convergence of the discretized solution to some summary statistics for the continuous function within each region. 

{With the above regularity conditions, we are ready to study the relationship between the minimizer of the working Poisson negative likelihood, $-\ell(\tilde{\balpha}, \bbeta)$, and the true parameters in the data generating mechanism in the low-dimensional setting without regularization. Proofs for Lemma~\ref{lem:target-param} and all other theoretical results are given in Appendix~\ref{app:proofs}.}

\begin{lemma}[Validity of working Poisson likelihood]
    For any set of $n$ locations $\boldsymbol s:=(s_1, \ldots, s_n) \in {\Omega_1\times \cdots\times \Omega_n}$, denote the vectorized form of the true intensity $\alpha^0(\cdot)$ as $\tilde\balpha^0(\boldsymbol s) = (\alpha^0(s_1), \ldots, \alpha^0(s_n))\in \mathbb R^n$,  
{and the vectorized form of the covariate surface as $X(\boldsymbol s)=(X(s_1)^\top,\ldots, X(s_n)^\top)^\top \in\mathbb R^{n\times p}$.
$X(\boldsymbol s)$ is allowed to differ from the discretized covariate matrix $\bX\in \mathbb R^{n\times p}$ used in model fitting.}
Let $\balphad(\boldsymbol s) := \tilde \balpha^0(\boldsymbol s) + {\left( X(\boldsymbol s) - \bX \right)\bbeta^0} + \phi(\boldsymbol s)$ for $\phi$ defined in Assumption~\ref{assump:regularity}. 

    Under \romannumeral1)--\romannumeral3) of Assumption~\ref{assump:regularity}, there exists $s_1^*\in\Omega_1, \ldots, s_n^*\in\Omega_n$ such that letting 
    \[\boldsymbol{\alpha}^\dagger(\boldsymbol s^*):=\left(\alpha^0(s_1^*)+{\left( X(s_1^*) - X_1 \right)\bbeta^0} + \phi(s_1^*),\ldots,\alpha^0(s_n^*)+{\left( X(s_n^*) - X_n \right)\bbeta^0}+\phi(s_n^*)\right)
    \]
    and $\boldsymbol{\beta}^\dagger:=\boldsymbol\beta^0$, we have
    \[
        -\nabla_{(\boldsymbol{\tilde\alpha}, \boldsymbol\beta)} \mathbb P_0\ell(\boldsymbol{\tilde\alpha}, \boldsymbol\beta) \big\vert_{(\boldsymbol{\alpha}^\dagger(\boldsymbol s^*), \boldsymbol{\beta}^\dagger)}
        = 0, 
    \]
    where $\mathbb P_0$ denotes {expectation under} the true data generating mechanism.
    \label{lem:target-param}
\end{lemma}

{We suppress the locations $\boldsymbol s^*$ from the expression of ${\balpha}^\dagger$ when it does not cause confusion, and} call $\boldsymbol\theta^\dagger:=\big({\boldsymbol{\alpha}^\dagger}^\top, {\boldsymbol{\beta}^\dagger}^\top\big)^\top$ defined in Lemma~\ref{lem:target-param} the \emph{target parameter}, since it is what the working Poisson likelihood (on population level) 
would lead us to find.
Lemma~\ref{lem:target-param} states that the target slope parameter $\boldsymbol\beta^\dagger$ associated with the working Poisson likelihood is equal to the true slope $\boldsymbol\beta^0$, even though the stochasticity in the intensity as well as the continuous 
nature of 
$\alpha^0(\cdot)$ and 
$X(\cdot)$ are ignored. Such mis-specification 
leads to a systematic bias in the target intercepts (compared to the discretized true baseline $\tilde\balpha^0$), determined only by the distribution of the errors at a finite set of locations, instead of the whole error random field. 

{When the penalty terms are involved, we would need the magnitude of penalty terms to scale appropriately relative to the degree-of-freedom of the parameters, in order to further study the behavior of the penalized estimator. We now introduce the assumption on the scaling and structure of penalty terms: } 

\begin{assumption}[{Penalty scaling and structure}] 
{Recalling the definition of ${\balpha}^\dagger(\boldsymbol s)$ in Lemma~\ref{lem:target-param}, further define
\[
G_n^\dagger := \gamma_n\nabla_\balpha R(\balphad;\mathcal G_n) =
        \begin{cases}
            \gamma_n B_n^\top S_\infty\left(\frac{ B_n\boldsymbol{\alpha^\dagger}}{\xi}\right), \text{ for smoothed } \ell_1 \text{ fusion penalty;}\\
            \gamma_n \tilde L_n\boldsymbol{{\alpha}}^{\dagger}, \text{ for } \ell_2 \text{ fusion penalty}.
        \end{cases}
\]
We assume}
\hfill
    \begin{itemize}
        \item[\romannumeral1)] $\tau_n = O_P\left(\sqrt{\frac{\log p}{n}}\right)$;
        \item[\romannumeral2)]  under the partition $\Omega = \Omega_1\cup\ldots\cup\Omega_n$, we have 
        \({n^{-1/t}}{\lVert G_n^\dagger\rVert_t} =
    O_P(\rho_{t,n})\)
        for $t\in\{1,2\}$ and sequences $\rho_{1,n},\rho_{2,n}$ varying with $n$.
        \item[\romannumeral3)] ${\left\lVert\frac{1}{n}\bX^\top G_n^\dagger\right\rVert_\infty = O_P\left(\sqrt{\frac{\log p}{n}}\right)}$. 
    \end{itemize}
    \label{assump:pen-scale}
\end{assumption}

The rate in Assumption~\ref{assump:pen-scale}\,\romannumeral1) is common in
high-dimensional estimation \citep{negahban2012unified,hastie2019statistical}.
Condition \romannumeral2) calibrates the fusion penalty strength to how well
the imposed smoothness structure matches the target intercept $\balpha^\dagger$.
The connected regions need not have truly similar target intercepts; rather, the condition allows a larger $\gamma_n$ when neighboring regions are truly similar
in the components underlying $\balpha^\dagger$ (namely, $\alpha^0(\cdot)$,
$\phi(\cdot)$, and the discretization residual $(X(s)-\bX)\bbeta^0$), so that
$\lVert B_n\boldsymbol\alpha^\dagger(\boldsymbol s)\rVert_1$ or
$\lVert\tilde L_n\boldsymbol\alpha^\dagger(\boldsymbol s)\rVert_2$ is small.
When this smoothness specification is non-informative, $\gamma_n$ must be smaller, yielding weaker fusion regularization. Condition \romannumeral3) states that the graph non-smoothness term, after projection onto the covariate space, is allowed to persist, but not exceeding the scale of the unavoidable $p$-dimensional sampling noise.

\begin{assumption}[Sparsity of $\boldsymbol\beta^0$]
    The sparsity level of the true slope {$q=\lVert\boldsymbol\beta^0\rVert_0$ satisfies $q = o\left(\sqrt{\frac{n}{\log p}}\right)$}.
    \label{assump:sparse}
\end{assumption}

\begin{assumption}[Graph fusion structure]
    Define ${\boldsymbol \eta}^\dagger:={\boldsymbol\alpha}^\dagger + \bX\bbeta^\dagger$, and denote the neighborhoods around ${\boldsymbol\alpha}^\dagger$ and  ${\boldsymbol\eta}^\dagger$ as
$
    \mathcal A_n(a_{\alpha,n})
    :=
    \left\{
        \btalpha\in\mathbb R^n:
        \left\|
            \btalpha
            -
            {\boldsymbol\alpha}^{\dagger}
        \right\|_2
        \le
        a_{\alpha,n}
    \right\}
$, and 
$\mathcal E_n(a_{\eta,n})
    :=
    \left\{
        \boldsymbol\eta\in\mathbb R^n:
        \left\|
            \boldsymbol\eta
            -
            \boldsymbol\eta^\dagger
        \right\|_2
        \le
        a_{\eta,n}
    \right\}$.
We require that $R(\tilde{\balpha};\mathcal G_n)$ is continuously differentiable on
\(\mathcal A_n(a_{\alpha,n})\), and that
${\nabla}R(\tilde{\balpha};\mathcal G_n)$ is locally Lipschitz on this set, which is satisfied by both the smoothed $\ell_1$ and $\ell_2$ fusion penalty.

For any
\(\btalpha\in\mathcal A_n(a_{\alpha,n})\), let
\(\bar H_n(\btalpha)\) denote the generalized Hessian \citep{hiriart1984generalized} satisfying
\[
    \gamma_n \nabla R(\tilde{\balpha};\mathcal G_n)
    -\gamma_n \nabla R({\balpha}^\dagger; \mathcal G_n) =
    \bar H_n(\btalpha)
    \left(
        \btalpha - {\boldsymbol\alpha}^{\dagger}
    \right).
\]
We assume that with probably converging to 1, 
\begin{itemize}
    \item[\romannumeral1)]
    the local linear-predictor perturbation is controlled:
    \[
        \sup_{\substack{
            \btalpha\in\mathcal A_n(a_{\alpha,n})\\
            \boldsymbol\eta\in\mathcal E_n(a_{\eta,n})
        }}
        \left\|
            \frac{1}{n}
            \bX^\top
            \bar H_n(\btalpha)
            \left(
                \boldsymbol\eta
                -
                \boldsymbol\eta^\dagger
            \right)
        \right\|_\infty
        =
        O_P\left(\sqrt{\frac{\log p}{n}}\right);
    \]
    \item[\romannumeral2)]
    the fusion compatibility condition holds: there exists
    a constant \(\kappa_F>0\) such that for
    \(\Delta\in\mathbb R^p\) and 
    \(\btalpha\in\mathcal A_n(a_{\alpha,n})\),
    \[
        \frac{1}{n}
        \Delta^\top
        \bX^\top
        \bar H_n(\btalpha)
        \bX
        \Delta
        \ge
        \frac{\kappa_F}{q}
        \|\Delta\|_1^2
        \tag{$\ell_1$ fusion compatibility}
    \]
    \[
        \frac{1}{n}
        \Delta^\top
        \bX^\top
        \bar H_n(\btalpha)
        \bX
        \Delta
        \ge
        {\kappa_F}
        \|\Delta\|_2^2
        \tag{$\ell_2$ fusion compatibility}
    \]
    \item[\romannumeral3)] \(\bar H_n(\tilde{\boldsymbol\alpha})\) is symmetric positive semidefinite on
\(\mathcal A_n(a_{\alpha,n})\). Furthermore,
\[
    \frac{1}{\sqrt n}
    \sup_{\boldsymbol\alpha\in\mathcal A_n(a_{\alpha,n})}
    \|
        \bar H_n(\boldsymbol\alpha)\bX
    \|_{\operatorname{op}}
    =
    O_P(\kappa_{2,n});
\]
and, only if a direct $\ell_1$ bound for
\(\boldsymbol\alpha^*-\boldsymbol\alpha^\dagger\) is required,
\[
    \frac{1}{n}
    \sup_{\boldsymbol\alpha\in\mathcal A_n(a_{\alpha,n})}
    \|
        \bar H_n(\boldsymbol\alpha)\bX
    \|_{1,1}
    =
    O_P(\kappa_{1,n}),
\]
for sequences $\kappa_{1,n}, \kappa_{2,n}$ varying with $n$, where
\(
    \|A\|_{1,1}
    :=
    \sup_{\|v\|_1=1}\|Av\|_1
    =
    \max_j\sum_i |A_{ij}|.
\)
\end{itemize}
\label{assump:graph-curv}
\end{assumption}

{
The use of generalized Hessian makes our methodology generalizable to fusion penalties that are not second-order continuously differentiable, such as the smoothed $\ell_1$ fusion penalty \citep{chen2012smoothing}. For $\ell_2$ fusion, the generalized Hessian reduces to the standard one, i.e., $\tilde L_n$.
Assumption~\ref{assump:graph-curv} controls the curvature of the graph fusion penalty in two ways. Condition \romannumeral1) requires that, in a small neighborhood around the target parameter, movements in the residual linear predictor would not lead to large perturbations to the projected score of $\bbeta$. In parallel, \romannumeral2) enforces sufficient local curvature induced by the graph fusion term and guarantees the identifiability of $\bbeta$. 
Under $\ell_2$ fusion, \romannumeral2) reduces to curvature  lower bound on $\bX^\top\tilde L_n\bX$ which is guaranteed if the design matrix $\bX$ is well-conditioned and the rates of $\gamma_n,\delta$ are selected appropriately.
}

\begin{assumption}[Bounded intensity]
\label{assump:bounded-cont-intensity}
The continuous unconditional intensity is uniformly bounded:
\(
    0<\psi
    \le
    \exp\left\{
        \alpha^0(s)
        +
        X(s)\boldsymbol\beta^\dagger
        +
        \phi(s)
    \right\}
    \le
    \Psi
    <\infty,
    \ s\in\Omega.
\)
\end{assumption}

\begin{assumption}[Design matrix]
    The design matrix $\boldsymbol X$ satisfies $\max_{i,j}|X_{ij}| \le R <\infty$ and $\lVert \bX\rVert_{\text{op}} = O_P(\sqrt{n})$.
\label{assump:design}
\end{assumption}

Consider, for the moment, the low-dimensional $\boldsymbol\beta^0$ without the $\ell_1$ sparsity penalty. With the assumptions introduced above, we now examine the minimizer of the combination of the loss function along with the fusion penalty, and investigate its relationship with the true baseline intensity $\alpha^0(\cdot)$ and regression parameters $\boldsymbol\beta^0$.

\begin{lemma}[Validity of PMLE with fusion penalty]
    Under Assumptions~\ref{assump:regularity}--\ref{assump:design}, denote
    \[
    {\boldsymbol\theta}^*
    =
    \left((\boldsymbol{\alpha}^*)^\top,(\boldsymbol{\beta}^*)^\top\right)^\top
    :=
    \argmin_{\boldsymbol{\tilde\alpha},\boldsymbol\beta}
    -\mathbb P_0\ell(\boldsymbol{\tilde\alpha},\boldsymbol\beta)
    +
    \gamma_n R(\boldsymbol{\tilde\alpha};\mathcal G_n).
    \]
    Then, for \(t=1,2\),
    \[
        \|\boldsymbol\beta^*-\boldsymbol\beta^\dagger\|_t
        =
        O_P\left(
            q^{1/t}\sqrt{\frac{\log p}{n}}
        \right),
    \]
    where only the rate for \(t=2\) requires the $\ell_1$ fusion compatibility condition in
    Assumption~\ref{assump:graph-curv}-\romannumeral2). Furthermore, for \(t=1,2\),
    \[
        n^{-1/t}
        \left\lVert
            \boldsymbol{\alpha}^*-\boldsymbol{\alpha}^\dagger
        \right\rVert_t
        =
        O_P\left(
            \rho_{t,n}
            +
            (1+\kappa_{t,n})
            q^{1/t}
            \sqrt{\frac{\log p}{n}}
        \right),
    \]
    where only the rate for \(t=2\) requires the second part of
    Assumption~\ref{assump:graph-curv}-\romannumeral3).
    \label{lem:target-param-fusion}
\end{lemma}


\begin{remark}
    A direct corollary of Lemma~\ref{lem:target-param-fusion} is that, when $\rho_{2,n}=o(1),
    \kappa_{2,n}
    \sqrt{\frac{q\log p}{n}}
    = o(1),
    \frac{q\log p}{n}=o(1),
    $
it holds that
\(
    n^{-1/2}
    \|
        \boldsymbol\alpha^*-\boldsymbol\alpha^\dagger
    \|_2
    =
    o_P(1).
\)
And when $
\rho_{1,n}=o(1),
    \kappa_{1,n}
    q\sqrt{\frac{\log p}{n}}
    =
    o(1),
    q\sqrt{\frac{\log p}{n}}
    =
    o(1)$,
we have
\(
    n^{-1}
    \|
        \boldsymbol\alpha^*-\boldsymbol\alpha^\dagger
    \|_1
    =
    o_P(1).
\)
\end{remark}

With the fusion penalty imposed, Lemma~\ref{lem:target-param-fusion} shows that
the low-covariate-dimensional optimization problem yields an close approximation $\bbeta^*$ relative to $\bbeta^0$. Under
Assumption~\ref{assump:pen-scale}, the penalty strength is calibrated to the
smoothness of the target intercept components: the baseline intensity
$\alpha^0(\cdot)$, error field $\phi(\cdot)$, and covariate-discretization
residual $(X(\cdot)-\bX)\bbeta^0$. 
When the imposed smoothness is less
consistent with the true mechanism, $\rho_{1,n}$ or $\rho_{2,n}$ dominates the
gap between $\balpha^*$ and $\balpha^\dagger$, leading to slower convergence
than the canonical rate. Conversely, when the structure is informative and the
total variation of
$\alpha^0(\cdot)+(X(\cdot)-\bX)\bbeta^0+\varepsilon(\cdot)$ is bounded,
$\rho_{1,n}$ and $\rho_{2,n}$ are small, so the penalty-induced gap between
$\balpha^*$ and $\balpha^\dagger$ vanishes pointwise.

{We next state the additional condition needed to establish high-dimensional consistency when the sparsity penalty is imposed on \(\bbeta\). }
{To this end, let $\mu_i^0:=\mathbb E_0Y_i=|\Omega_i|
    \exp\{\alpha_i^\dagger+X_i^\top\bbeta^0\}$ following Lemma~\ref{lem:target-param}, and write
\(Z_i:=Y_i-\mu_i^0\) and 
\(    \boldsymbol Z:=(Z_1,\ldots,Z_n)^\top.
\)
Let
\(
    S:=\operatorname{supp}(\bbeta^0)
\) be the support of $\bbeta^0$ with $ |S|=q.$}

\begin{assumption}[Localized empirical-process bound]
\label{assump:centered-score}

{
Let \(a_n\) be a sequence 
such that $qa_n=o(1)$ and $\tau_n\asymp a_n$.
For an increment along the direction of $\btheta=\left({\balpha}^\top, \bbeta^\top\right)^\top$, written as
\(
    \Delta=(\Delta_\alpha^\top,\Delta_\beta^\top)^\top \in\mathbb R^{n+p},
\)
let
\(
    \Delta_\eta:=\Delta_\alpha+\bX\Delta_\beta.
\)
}
Moreover, for constants \(c_1,c_2>0\), define
\[
    \mathcal C_n(c_1,c_2)
    :=
    \left\{
        \Delta=(\Delta_\alpha^\top,\Delta_\beta^\top)^\top:
        \|\Delta_{\beta,S^C}\|_1
        \le
        c_1\|\Delta_{\beta,S}\|_1+c_2qa_n
    \right\}.
\]
We assume there exists a constant \(C_Z<\infty\) such that, with probability converging to 1,
\[
    \left|
        \frac1n\boldsymbol Z^\top\Delta_\eta
    \right|
    \le
    C_Z
    \frac{\|\Delta_\eta\|_2}{\sqrt n}
    \sqrt q\,a_n
    +
    C_Zqa_n^2
\]
holds uniformly for all $\Delta\in\mathcal C_n(c_1,c_2)$.
\end{assumption}

Assumption~\ref{assump:centered-score} or similar variants are commonly imposed in high-dimensional M-estimation \citep{buhlmann2011statistics}. It is weaker than assuming a global empirical-process bound over all
\((n+p)\)-dimensional directions, and is restricted to the sparsity cone arising from the \(\ell_1\) penalty. Under sub-exponential or Bernstein-type concentration \citep{wainwright2019high} of the centered counts \(Z_i=Y_i-\mu_i^0\), together with a localized covering-number bound for the feasible linear-predictor increments, this condition holds with the standard rate
\(
    a_n\asymp\sqrt{\frac{\log p}{n}}.
\)
For heavier-tailed Cox-processes, the same condition may hold with a slower rate \(a_n\) under finite-moment and weak-dependence assumptions. 

Note that the standard restricted strong convexity condition \citep{negahban2012unified} is not separately required here; rather, it is derived (see Appendix~\ref{app:proofs}) from the existing bounded-intensity assumption, the asymptotic regime, and the graph-fusion structure assumption.
We now establish our consistency result.

\begin{theorem}[Consistency of penalized PMLE]
    Let
\(
    \hat{\btheta}
    =
    (\hat{\balpha}^{\top},\hat{\bbeta}^{\top})^\top
\)
be the solution to \eqref{equ:obj-fun}, and recall the definition of $\btheta^*$ in Lemma~\ref{lem:target-param-fusion}.
Define the corresponding linear predictors (with the discretized covariates)
\(
    \hat{\boldsymbol\eta}
    :=
    \hat{\balpha}+\bX\hat{\bbeta},
    \boldsymbol\eta^*
    :=
    \balphas+\bX\bbeta^*.
\)
Under Assumptions~\ref{assump:regularity}--\ref{assump:centered-score}, we have
\[
    \|\hat{\bbeta}-\bbeta^*\|_t
    =
    O_P(q^{1/t}a_n),
    \qquad
    n^{-1/t}
    \|\hat{\boldsymbol\eta}-\boldsymbol\eta^*\|_t
    =
    O_P(\sqrt q\,a_n)
\]
for $t\in\{1,2\}$, where $a_n$ is defined as in Assumption~\ref{assump:centered-score}.

Following Lemma~\ref{lem:target-param-fusion} and by the triangular inequality, this immediately implies 
\[
    \left\|\hat{\bbeta}-\bbeta^0\right\|_t
    =
    O_P\left[q^{1/t}\left(a_n + \sqrt{\frac{\log p}{n}}\right)\right] \text{ for } t \in \{1,2\}.
\]
\label{thm:consistency}
\end{theorem}
\begin{remark}
    Under the canonical Bernstein-type concentration \citep{wainwright2019high}, it holds that $a_n\asymp \sqrt{\frac{\log p}{n}}$, and hence the convergence rates in Theorem~\ref{thm:consistency} reduce to the classical $ O_P\left(
        q\sqrt{\frac{\log p}{n}}
    \right)$ rate for $\ell_1$ error bound, and $O_P\left(
        \sqrt{\frac{q\log p}{n}}
    \right)$ for $\ell_2$.
\end{remark}

\section{{Statistical Inference}} 
\label{sec:inference}



In this section, we construct confidence intervals for each
$\beta^0_j$, $j=1,\ldots,p$, and establish asymptotic normality while
accounting for the doubly stochastic variation not captured by the PMLE
estimating equation. The result extends directly to linear contrasts of
multiple $\beta$'s.
It is known that penalized $M$-estimators are generally biased
\citep{voorman2014inference} and their uncertainty is difficult to
characterize analytically \citep{zhao2021defense}. Here, we adopt a de-biasing approach based on \citet{javanmard2014confidence}. Relative to their original procedure, our extension allows non-Gaussian models and accounts for the extra randomness from the error random field using a conservative sandwich covariance estimator.



A general de-biasied estimator takes the form $\hat{\boldsymbol b} = \hat{\boldsymbol\beta} + n^{-1} M\nabla_{\boldsymbol\beta} \ell(\hat{\boldsymbol\alpha}, \hat{\boldsymbol\beta})$, where the choice matrix of $M$ determines how well the bias and variance are controlled by the inference procedure. In our setting, such an estimator is given by
\[
    \hat{\boldsymbol b} = \hat{\boldsymbol\beta} + \frac{1}{n} M\boldsymbol X^\top \left[\boldsymbol Y - \boldsymbol B\odot \exp\left(\hat{\boldsymbol\alpha} + \boldsymbol X\hat{\boldsymbol\beta}\right)\right],
\]
where 
$\boldsymbol B=(|\Omega_1|,\ldots,|\Omega_n|)$, and $\odot$ indicates element-wise multiplication. 
Our choice of $M$ is based on two quantities, the empirical Hessian of the {negative} Poisson log-likelihood, 
\(
    \hat{\boldsymbol H}= -\frac{1}{n}\sum_{i=1}^n \nabla^2_\beta\ell(\hat{\balpha}, \hat\bbeta; x_i, y_i),
\)
and an estimated covariance $\hat{\boldsymbol{\Sigma}}$ of the gradient $\nabla_{\boldsymbol\beta} \ell(\balpha^\dagger, \bbeta^0)$. 
Note that using a plug-in estimate $\hat{\boldsymbol H}$ to derive $\hat{\boldsymbol{\Sigma}}$ would underestimate the variability due to the stochasticity of baseline intensity. Instead, we use a conservative covariance estimate 
\begin{equation}
    \hat{\boldsymbol\Sigma}:= \frac{2}{n}\sum_{i=1}^n X_i^\top X_i\left[ \left(Y_i - \lvert\Omega_i\rvert P_i \exp(\hat\alpha_i + X_i\hat\bbeta)\right)^2 + \left( \lvert\Omega_i\rvert P_i \exp(\hat\alpha_i + X_i\hat\bbeta) - \bar\mu\right)^2 \right],
    \label{equ:Sig-hat}
\end{equation}
where $\bar\mu := n^{-1} \sum_i \lvert\Omega_i\rvert P_i \exp(\hat\alpha_i + X_i\hat\bbeta)$. The first term in (\ref{equ:Sig-hat}), without the multiplier 2, is a natural estimator for Poisson (not doubly-stochastic) data, and the added terms capture the additional stochasticity in the latent intensity. 

Finally, $M$ is defined such that its $j$th row, $m_j$ is the solution of
\begin{equation}
    \min_{m} m\hat\bSigma m^\top \quad
    \text{s.t.} \quad \lVert \hat{\boldsymbol H} m ^\top - e_j \rVert_\infty \le {\zeta}
    \label{equ:m-debias}
\end{equation}
with $e_j$ being the vector with one at the $j$th entry and zero everywhere else, and $\zeta$ being a small tolerance parameter. Extending \citet{javanmard2014confidence}, the optimization problem (\ref{equ:m-debias}) aims to control two quantities: $\max_{i,j} |(\hat{\boldsymbol H} M - I_p)_{ij}|$ corresponding to the non-Gaussianity and bias of $\hat{\boldsymbol b}$, and $(M\hat\bSigma M)_{ii}$ relating to the variance of $\hat{\boldsymbol b}$. However, (\ref{equ:m-debias}) differs from the original optimization problem proposed by \citet{javanmard2014confidence} in that the bias and variance are captured separately by $\hat\bSigma$ and $\hat{\boldsymbol H}$ in our setting. This is expected since the first-order properties of the penalized PMLE are determined by the Poisson log-likelihood, while the doubly-stochastic nature of the true process needs to be accounted for when characterizing second-order properties. 

{The inference procedure would require some additional assumptions beyond those stated in Theorem~\ref{thm:consistency}. 
Let \(\mathcal J\subseteq\{1,\ldots,p\}\) denote the
coordinates for which inference is conducted. For each \(j\in\mathcal J\), let
\(m_j\in\mathbb R^p\) denote the population-level counterpart of the \(j\)th row
of \(M\). Let
\(
    \eta_i^* := \alpha_i^* + X_i^\top\bbeta^*, 
    \mu_i^* := |\Omega_i|\exp(\eta_i^*),
    Z_i^* := Y_i-\mu_i^*
\)
and define
\[
    \sigma_j^2
    :=
    \frac1n
    \operatorname{Var}_0
    \left(
        m_j^\top\bX^\top\boldsymbol Z^*
    \right)
    =
    \frac1n
    \sum_{i,k=1}^n
    (m_j^\top X_i)(m_j^\top X_k)
    \operatorname{Cov}_0(Z_i^*,Z_k^*),
\]
where \(\boldsymbol Z^*=(Z_1^*,\ldots,Z_n^*)^\top\), and
\(\operatorname{Var}_0(\cdot)\) and \(\operatorname{Cov}_0(\cdot,\cdot)\)
are taken under the true data-generating mechanism.

The additional conditions, stated precisely in Appendix~\ref{app:inference},
strengthen the consistency requirements in the directions needed for inference.
In particular, they control the empirical-process and debiasing approximation
errors at the \(n^{-1/2}\) scale, require local stability of the Hessian
\(
    H(\boldsymbol\eta)
    :=
    \frac1n \bX^\top
    \operatorname{diag}\{|\Omega_i|\exp(\eta_i)\}_{i=1}^n
    \bX
\)
around \(\boldsymbol\eta^*=(\eta_1^*,\ldots,\eta_n^*)^\top\), and ensure that
the remaining nuisance-intercept and graph-fusion contributions are
asymptotically negligible after debiasing. They also impose nondegeneracy of the limiting variance \(\sigma_j^2\).


\begin{theorem}[Asymptotic normality]
Under Assumptions~\ref{assump:regularity}--\ref{assump:centered-score} and the additional regularity conditions stated under Assumption~\ref{assump:inference} in Appendix~\ref{app:inference}, it holds that
\[
    \frac{
        \sqrt n(\hat b_j-\beta_j^0)
    }{
        \sigma_j
    }
    \xrightarrow{\mathrm d}
    N(0,1).
\]
\label{thm:asym-normal}
\end{theorem}
We show in Appendix~\ref{app:inference} that $[M\hat\Sigma M^\top]_{jj}$ as defined in (\ref{equ:Sig-hat}) serves as a conservative estimator of  $\sigma_j$. 
Although the inference procedure is distribution-free with respect to the error field, a known error distribution could improve efficiency by allowing the population covariance
\(\mathbb E_0[
\nabla_{\boldsymbol\beta}\ell(\hat{\boldsymbol\alpha},\hat{\boldsymbol\beta})
\nabla_{\boldsymbol\beta}\ell(\hat{\boldsymbol\alpha},\hat{\boldsymbol\beta})^\top]\)
to be expressed using estimated variance parameters.

\section{Simulations}
\label{sec:sim}

\subsection{{Estimation and Inference Performance}}

We compare penalized PMLE with alternative approaches for LGCP using 100 simulated
replicates on $\Omega=[0,m]\times[0,m]$, partitioned into $n=m^2$ unit-square
cells. Following \eqref{equ:intensity}, the baseline intensity is
\(\alpha^0(s)=\frac{1}{4m}\sqrt{s_1^2+s_2^2}\) for
\((s_1,s_2)\in\Omega\). The random error \(\varepsilon(\cdot)\) consists of a
spatially structured component and an unstructured component. The structured
component is generated from a zero-mean Gaussian random field with exponential
covariance and range parameter \(0.2m\); the unstructured component is generated
on a fine \(60\times 60\) grid, is constant on each small cell, and has
independent Gaussian values with variances drawn from an inverse Gamma
distribution with shape 2 and rate 1 to induce non-stationarity. 
Though a Gaussian random field is continuous, it is typically discretized and simulated on fine grids in practice, as is our case for $\alpha^0(s)$ and $\varepsilon(s)$.
Each entry of
the \(p\)-dimensional covariate \(X\) is drawn from Uniform\([-0.5,0.5]\), with
locations in the same cell sharing covariate values, and \(P(s)=2\). We consider
two settings: (i) \(p=10\), with \(\beta_1=\beta_2=-1\),
\(\beta_3=\beta_4=1\), and \(\beta_5=\cdots=\beta_{10}=0\); and (ii) \(p=100\),
with \(\beta_1=\cdots=\beta_5=-1\), \(\beta_6=\cdots=\beta_{10}=1\), and all
remaining entries zero. We investigate \(n=5^2,10^2,20^2,30^2\), and use an
unweighted graph \(\mathcal G_n\) connecting horizontally or vertically adjacent
cells.

We compare PMLE with \(\ell_1\) and \(\ell_2\) fusion penalties, where
\(\gamma_n\) and \(\tau_n\) are jointly selected by 5-fold cross-validation to
minimize prediction MSE, against three benchmarks: \romannumeral1) LGCP with Gaussian random errors and exponential covariance, fitted by
    \texttt{RStan} using 1000 posterior MCMC samples, with Normal\((0,10)\)
    priors for slopes and truncated Normal\((0,5)\) priors for covariance
    parameters; \romannumeral2) \texttt{RStan} using 1000 posterior MCMC samples, with Normal\((0,10)\)
    priors for slopes and truncated Normal\((0,5)\) priors for covariance
    parameters; \romannumeral3) Scampr, the variational approximation and reduced-rank interpolation
    method of \citet{dovers2023fast} implemented in the \texttt{scampr}
    \texttt{R} package.

Figure~\ref{fig:comp-time} shows the average computation time. Scampr, PMLE and
INLA all scale reasonably well as dimension and sample size increase. PMLE and
Scampr are faster than INLA in both settings, while MCMC sampling via
\texttt{RStan} is time-consuming for large \(p\) and/or \(n\); therefore,
\(n=30^2\) is not examined for \texttt{RStan}.

\begin{figure}
    \centering
    \includegraphics[width = 12cm]{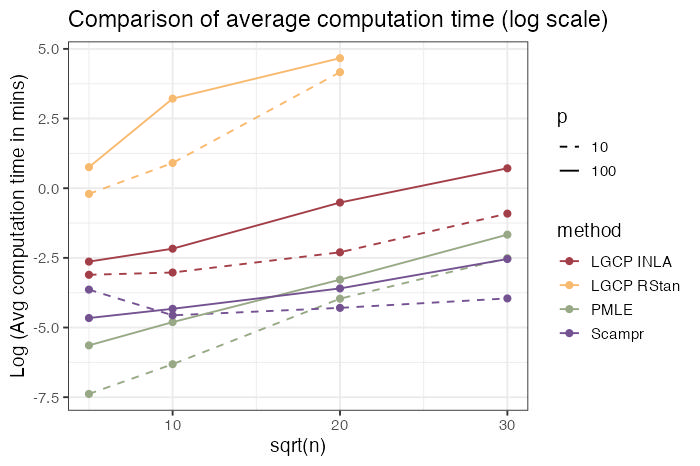}
    \caption{Average computation time for a single replicate of data in minutes,
    plotted on log scale, over 100 replicates for penalized PMLE, Scampr and
    Bayesian LGCP models run via \texttt{RStan} and \texttt{R-INLA}.}
    \label{fig:comp-time}
\end{figure}

Table~\ref{tab:sim-mse} compares the average entry-wise estimation MSE under
\(n=30^2\). All models achieve comparable estimation performance, with penalized
PMLE demonstrating higher accuracy in the high-dimensional setting.

\begin{table}[ht]
\centering
\begin{tabular}{@{}lllll@{}}
\toprule
        & PMLE L1 & PMLE L2 & INLA   & Scampr \\ \midrule
$p=10$  & 0.0678  & 0.0744  & 0.0537 & 0.0799 \\
$p=100$ & 0.0184  & 0.0184  & 0.0218 & 0.0260 \\ \bottomrule
\end{tabular}
\caption{Average entry-wise estimation MSE under \(n=30^2\); LGCP with
\texttt{RStan} is omitted due to intensive computation.}
\label{tab:sim-mse}
\end{table}

\begin{figure}
    \centering
    \includegraphics[width=16cm]{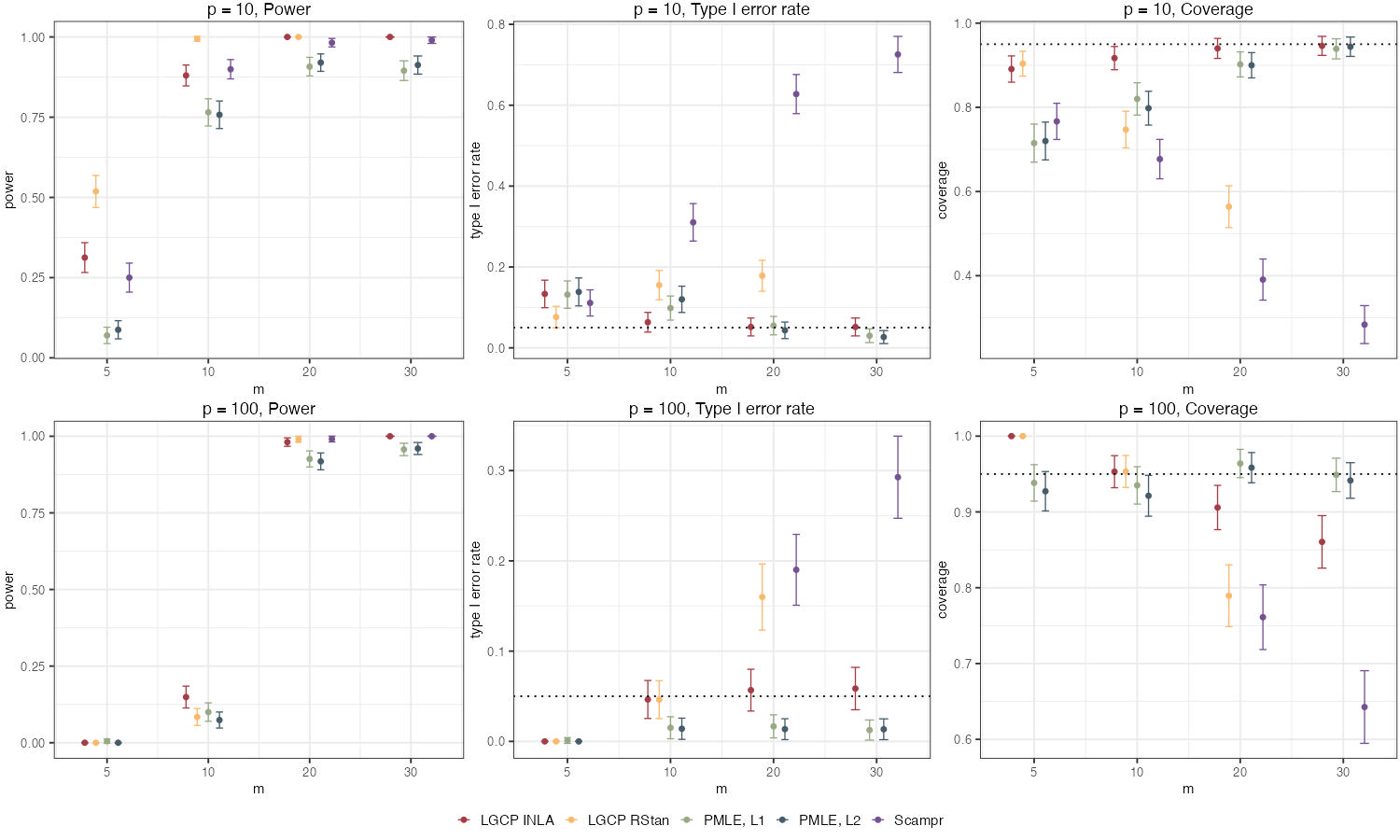}
    \caption{Comparison of coverage, type I error rate and power for penalized
    PMLE, Scampr and Bayesian LGCP methods, with standard error bars.}
    \label{fig:sim_inf}
\end{figure}

Figure~\ref{fig:sim_inf} compares coverage, type I error rate and power,
averaged across the relevant entries of \(\bbeta\). In low dimensions,
\texttt{R-INLA} performs well, with power approaching 1, controlled type I error
and valid \(95\%\) coverage. Penalized PMLE achieves similar accuracy but
requires more samples, due to its over-parameterized structure and its lack
of reliance on a parametric distribution for \(\varepsilon(\cdot)\).
\texttt{RStan} does not control type I error or achieve proper coverage with the
given amount of data and MCMC samples.

In high dimensions, \texttt{R-INLA} no longer consistently achieve nominal
coverage or type I error control. This may come from violation
of the constant-baseline-risk assumption for the RW2D model, the presence of non-stationarity, as well as inadequate degree-of-freedom control in the slope priors. 

Scampr fails for \(m=5,10\) with \(p=100\),
and otherwise shows inflated type I error and lower coverage; together with the
MSEs in Table~\ref{tab:sim-mse}, this suggests variance underestimation rather
than poor point estimation. In contrast, penalized PMLE controls type I error
within 0.05 and maintains reasonable power, although it is conservative due to its weaker distributional assumptions. Additional results with a vanilla, correctly-specified stationary LGCP are given in Appendix~\ref{app:addn_rslts}, where penalized PMLE demonstrates conservativeness, but still achieves comparable power as INLA with sufficient sample size.

\subsection{Sensitivity Analysis}

We examine the sensitivity of our approach to the graph specification used in the fusion penalty. To this end, we compare the estimated slope parameters under three graph structures: \romannumeral1) the
unweighted horizontal/vertical adjacency graph used above; \romannumeral2) a graph that
additionally connects diagonal neighbors; and \romannumeral3) the original graph with
horizontal edges assigned half the weight of vertical edges.

Figure~\ref{fig:sim-graph-sensitivity} shows the entry-wise estimates under \(p=100\) and \(m=10^2\). The estimates are similar across graph specifications,
indicating robustness of penalized PMLE against alternative connectivity and weighting choices.
The nonzero entries are attenuated, as expected under the sparsity penalty, but this is expected to improve after de-biasing, and the pattern is consistent across graph specifications. 

Additional sensitivity and numerical analyses are presented in Appendix~\ref{app:addn_rslts}. Specifically, Figure~\ref{fig:sim-cv-surface} shows the cross-validation surfaces for the tuning parameters $(\gamma,\tau)$ under different graph specifications. Figures~\ref{fig:sim-base-int-gap} and~\ref{fig:sim-sparsity} further examine baseline intensity estimation and degree-of-freedom control under $\ell_1$ and $\ell_2$ fusion penalties.

\begin{figure}[ht]
    \centering
    \includegraphics[width=17cm]{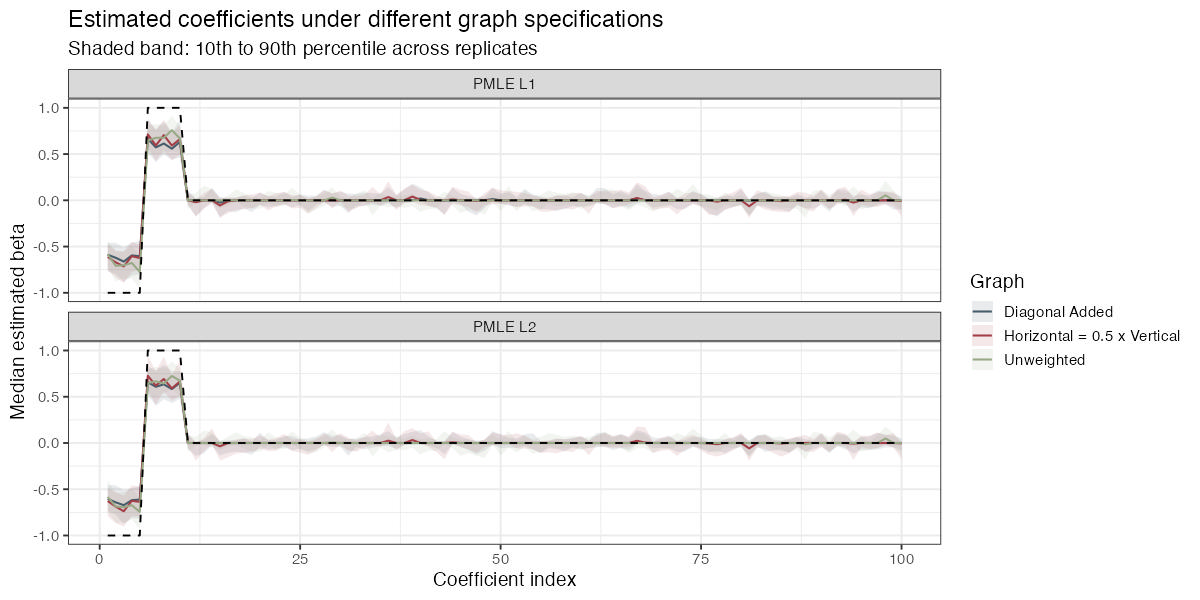}
    \caption{Entry-wise estimated values of \(\beta\) under different graph
    specifications, in the \(p=100, m=10^2\) scenario. The \(10\)th and \(90\)th
    percentiles across replicates are indicated as confidence bands. The dashed
    line represents the true values.}
    \label{fig:sim-graph-sensitivity}
\end{figure}
\section{Application: Seattle Crime Data}
\label{sec:data}

We analyze the Seattle crime data\footnote{\url{https://www.seattle.gov/police/information-and-data/crime-dashboard}}
to demonstrate the performance of our approach against several alternatives. We focus on crimes against persons reported to the Seattle Police Department in Spring 2021, April 1 through June 30. Crime cases are recorded as point incidents, with blurred locations, over the Seattle map. We aggregate incident counts to census tracts, the finest resolution at which covariates are available, and use tract population size as the offset. Covariates are obtained from King County GIS Open Data\footnote{\url{https://www.kingcounty.gov/services/gis/GISData.aspx}} and include demographic and socioeconomic variables, including age and race/ethnicity distributions, median household income, college-education rate, and medical-insurance rate; public-facility counts, including hospitals, transit stops, fire stations, police stations, food facilities, schools, solid-waste facilities, and farmers' markets; and environmental variables, including tract area and proportions of medium and high basins.

We intentionally choose a wide range of covariates, including those not directly known as good predictors of crimes, so that the analysis includes both strongly and weakly informative predictors. Covariates are all summarized by census tract, and for those characterized by proportions of different groups, such as age, race/ethnicity, and medium/high basins, we omit one category as the reference level and adopt the additive log-ratio transformation \citep{aitchison1982statistical} to alleviate spurious correlation in compositional data. The spatial domain is modeled as an unweighted graph, where two regions are connected if they share a common border.

We compare penalized PMLE with $\ell_1$ and $\ell_2$ fusion penalties with Scampr and three Bayesian models implemented in \texttt{INLA}: BYM2 \citep{riebler2016intuitive}, LGCP with independent Gaussian errors, and LGCP with an exponential-covariance Gaussian random field. The PMLE tuning parameters $\gamma_n$ and $\tau_n$ are jointly selected via cross-validation as in Section~\ref{sec:sim}. The default penalized-complexity priors \citep{simpson2017penalising} in the \texttt{R-INLA} package are used for the variance, range, and mixing parameters.

\begin{figure}
    \centering
    \includegraphics[width=16.5cm]{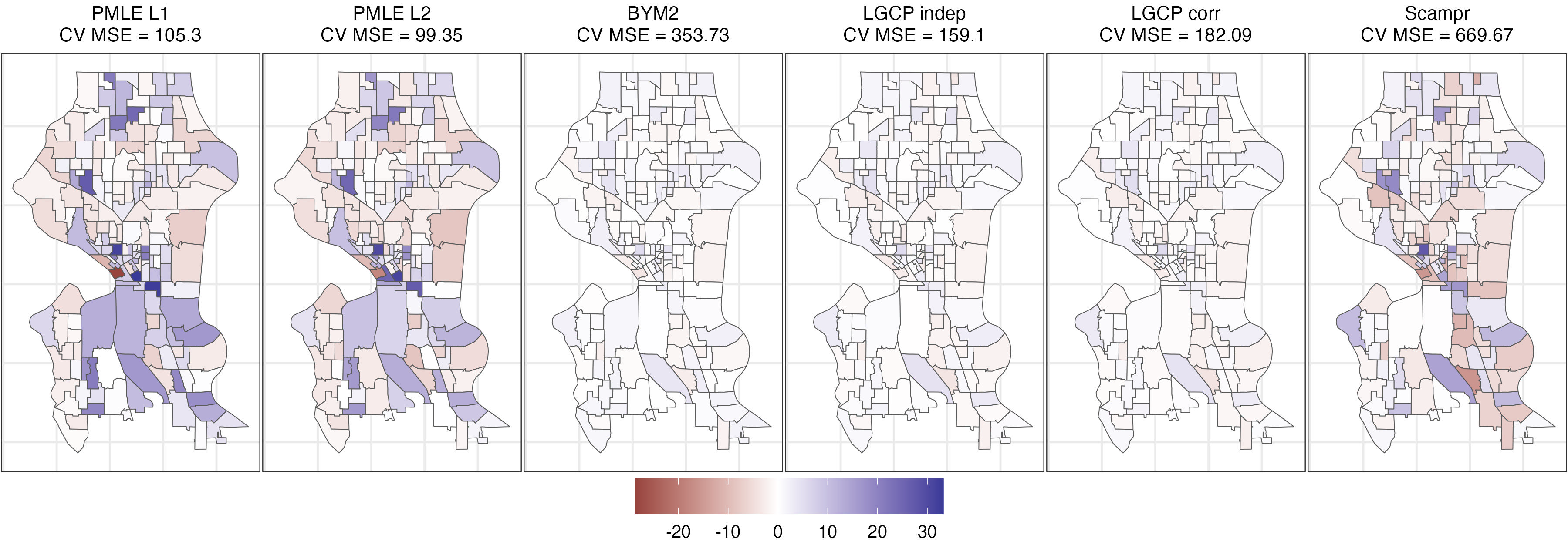}
    \caption{Residuals from each model, with cross-validated MSEs reported in the titles. The large MSEs for BYM2 and Scampr were partly driven by large variability in prediction errors among the cross-validation folds. For reference, their median SEs are 122.09 for BYM2 and 283.92 for Scampr.}
    \label{fig:data-res}
\end{figure}

We evaluate predictive performance using 5-fold cross-validation, with prediction MSE as the primary criterion. Other predictive scores, such as the conditional predictive ordinate \citep{gelfand1994bayesian}, are also useful, but are most natural for Bayesian models and less directly comparable across all methods considered here; therefore, we use MSE as the common criterion across methods. Figure~\ref{fig:data-res} displays full-data residual maps for each fitted model, with the corresponding cross-validated prediction MSE reported in each panel title. The residual maps provide in-sample diagnostics, whereas cross-validated MSE measures out-of-sample predictive accuracy and reflects the overall bias-variance trade-off.

The residual maps show that the Bayesian LGCP models and Scampr produce smaller full-data residual magnitudes than penalized PMLE, suggesting closer in-sample adaptation to local spatial variation. However, this does not translate into better predictive generalization: PMLE achieves the lowest prediction MSE among the methods considered. In particular, the Bayesian LGCP models appear to fit local spatial variation more aggressively, yielding lower full-data residuals but higher held-out prediction error, a pattern suggesting overfitting. Scampr shows a similar disconnect between its full-data residual map and its substantially larger prediction MSE. These results suggest that, in this application, the explicit penalization in PMLE provides a more favorable bias-variance trade-off for prediction.

\begin{figure}
    \centering
    \includegraphics[width=17cm]{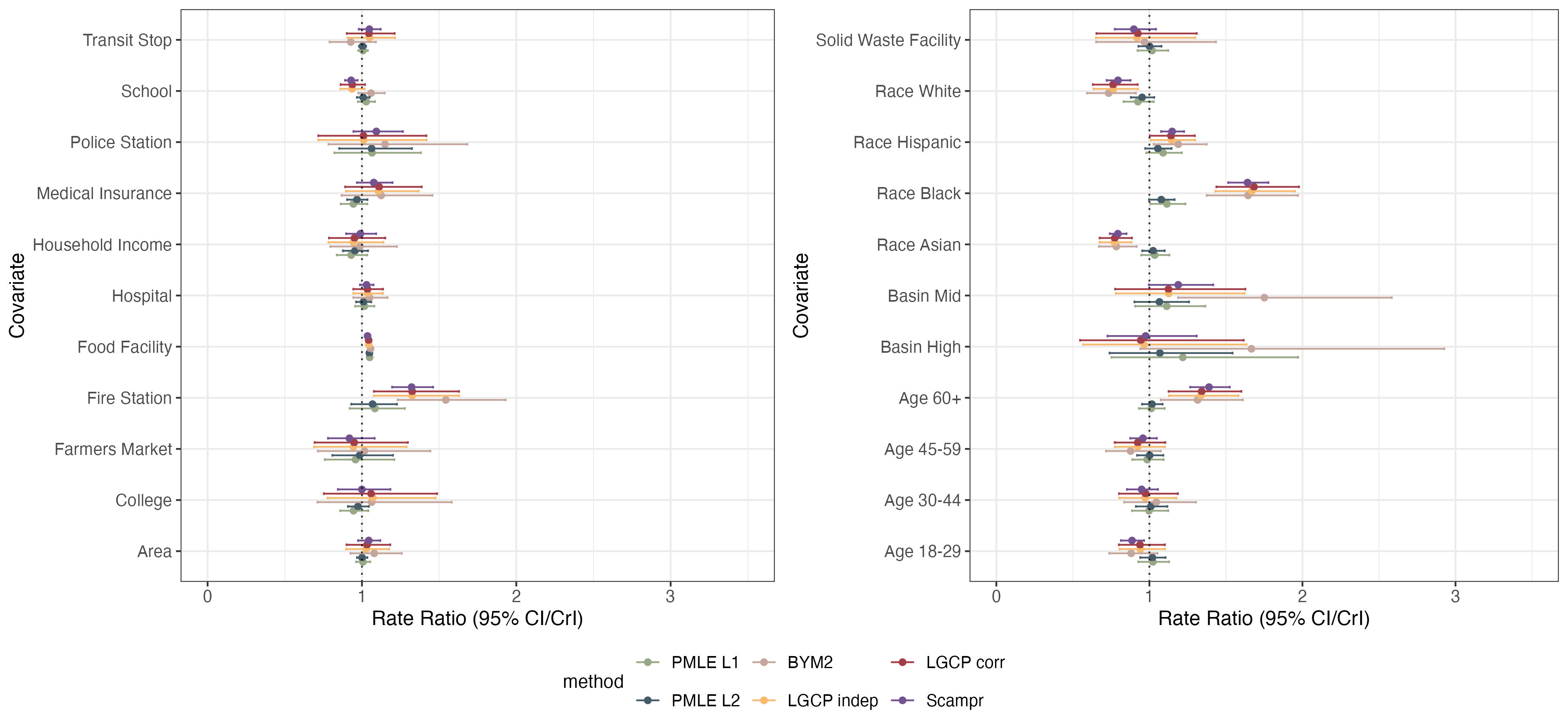}
    \caption{Estimated rate ratios with error bars indicating $95\%$ confidence/credible intervals.}
    \label{fig:crime-ests}
\end{figure}

Figure~\ref{fig:crime-ests} presents rate-ratio estimates with $95\%$ confidence/credible intervals. Across methods, race/ethnicity and the number of food facilities are consistently associated with crime incidence, with the race/ethnicity finding aligning with prior studies on housing inequalities and differential crime exposure by race \citep{uehara1994race, krysan2008does, lodge2021estimating}. PMLE and Scampr generally yield narrower intervals than the Bayesian methods. When estimates differ across methods, PMLE often gives intermediate values, as seen for transit stops and schools. Some additional associations identified by the other methods, including medium basin and the proportion of senior residents, are not clearly supported by existing studies or common knowledge. These findings are consistent with Section~\ref{sec:sim}, where PMLE showed better type-I error control without substantial loss of power.

A common concern in spatial data analysis is the effect of \emph{spatial confounding} \citep{reich2006effects, paciorek2010importance}. To assess the sensitivity of each method to spatially structured noise and increased covariate dimensionality, we repeat the analysis after adding 50 synthetic noise variables to the original covariates; see Appendix~\ref{app:addn_rslts} for details.
Overall, PMLE under both $\ell_1$ and $\ell_2$ fusion remains stable after the noise variables are introduced. As shown in Figure~\ref{fig:data-par-diff}, the estimated effects for the original covariates are similar before and after adding the synthetic noise variables, with tight confidence intervals. In contrast, the Bayesian methods and Scampr show greater sensitivity, reflected by wider CI/CrIs for some covariates and marginally significant sign changes in some cases. Figure~\ref{fig:data-noise-est} further shows that PMLE estimates near-zero effects for the synthetic noise variables, whereas the Bayesian methods and Scampr falsely identify several noise variables as statistically significant.
\section{Discussion}
\label{sec:discussion}

We proposed a computationally efficient semiparametric approach for estimating
and conducting inference on fixed covariate effects in doubly stochastic point
processes. The main contribution is to establish that a discretized Poisson working
model, although misspecified for the stochastic intensity, can still yield
consistent estimation and valid inference for the regression parameters with the
realized intensity represented through a flexible high-dimensional intercept.
This allows flexible bias-variance trade-off tuning, and avoids requiring a parametric model or strong second-order conditions for the latent intensity or covariance
structure. In our framework, region-specific intercepts with a fusion penalty
absorb the nonparametric baseline and realized latent variation, while a sparsity
penalty is used for the fixed covariate effects. The imposed smoothness structure
serves as regularization rather than a stringent requirement for the latent
field. We further account for the extra stochasticity from the doubly stochastic
process through robust covariance estimators.

The current approach does not
directly provide predictions beyond the aggregated level; as a potential extension, prediction at specific
locations could be pursued through a two-step procedure that smooths the
estimated intensity while plugging in the estimated covariate effects. In addition,
graph-denoising approximations may further reduce computation. For example, a
sparse approximation to the edge incidence matrix \(B_n\), as in
\citet{padilla2017dfs}, or an approximation to the graph Laplacian \(L_n\), as in
\citet{sadhanala2016graph}, could reduce the computational burden in large-scale
settings, while theoretical guarantees remain an interesting topic of future research.

Several practical questions also remain open. The debiasing threshold \(\zeta\)
in Equation~\ref{equ:m-debias} controls the trade-off between type I error and
power, and more principled guidance for choosing this parameter would be useful,
especially in finite samples. Similarly, prediction and parameter tuning for
graphical or spatial models remain theoretically delicate because naive
cross-validation may be affected by spatial dependence. Future work could
therefore study tuning strategies that either justify sample splitting under
dependence \citep{rabinowicz2022cross} or avoid it altogether. Finally, formal
goodness-of-fit tests and diagnostic procedures would further strengthen the
practical use of this semiparametric framework, which are not immediately
available from the classical likelihood-based theory.

\bibliographystyle{rss}
\bibliography{ref}

\newpage
\clearpage
\appendix

\pagenumbering{arabic}
\setcounter{page}{1}
\setcounter{figure}{0}
\setcounter{table}{0}
\renewcommand{\thefigure}{S\arabic{figure}}
\renewcommand{\thetable}{S\arabic{table}}

\begin{landscape}

\begin{center}
{\large\bfseries APPENDIX}
\end{center}

\section{Summary of Related Methods}
\label{app:lit-table}

\renewcommand{\arraystretch}{1.2}

\begin{table}[ht]
\centering
\scriptsize
\begin{tabular}{
p{4.2cm}
p{4.5cm}
p{4.8cm}
p{3.5cm}
p{4.5cm}
}

\toprule
\textbf{Approach}
& \textbf{Primary inferential target}
& \textbf{Treatment of spatial dependence}
& \textbf{Second-order specification required?}
& \textbf{High-dimensional inference supported?} \\
\midrule

Bayesian LGCP (MCMC, HMC) 
\citep{moller1998log,diggle2013spatial}
& Latent intensity surface; regression and covariance estimation with model-based uncertainty
& Explicit latent Gaussian field integrated via simulation
& Yes (parametric covariance model for latent field)
& Typically no (fixed or low-dimensional covariates) \\

INLA--SPDE / GMRF 
\citep{rue2009approximate,lindgren2011explicit}
& Latent field recovery and regression with approximate posterior inference
& Sparse GMRF approximation of Gaussian random field via SPDE construction
& Yes (SPDE structure and covariance family required)
& Limited (primarily low-dimensional regression settings) \\

Reduced-rank / variational LGCP ML 
\citep{dovers2023fast}
& Regression and covariance estimation via approximate marginal likelihood
& Latent field approximated by basis expansion or variational family
& Yes (basis choice and covariance specification)
& Limited theoretical development for high-dimensional inference \\

Aggregated-data LGCP 
\citep{li2012log}
& Continuous-space risk surface and regression from areal counts
& Latent Gaussian field combined with aggregation operator
& Yes (latent covariance specification)
& No explicit high-dimensional inference framework \\

Poisson score as estimating function 
\citep{schoenberg2005consistent,guan2007thinned}
& Regression parameter estimation via first-order inference
& Uses unbiased first-order moment identities under clustering
& No parametric second-order model required
& Developed mainly for low-dimensional parameter settings \\

Composite / Palm likelihood 
\citep{guan2006composite,baddeley2017local}
& Regression and/or interaction parameter estimation
& Lower-order (pairwise/Palm) likelihood components
& Typically yes (pair correlation; stationarity or reweighted stationarity assumptions)
& Generally low-dimensional parameter regimes \\

Quasi-likelihood / weighted estimating equations 
\citep{guan2010weighted,guan2015quasi,deng2017second}
& Efficient regression inference under clustering
& Incorporates second-order structure into estimating equations
& Yes (pair correlation or related integral structure required)
& Not developed for diverging-dimensional covariates \\

Regularized spatial point process regression 
\citep{choiruddin2018convex,choiruddin2023adaptive}
& Variable selection and penalized regression estimation
& First-order (Poisson/logistic-type) criteria under dependence
& No full latent-field modeling
& Consistency and sparsity theory available; post-selection inference limited \\

\textbf{Proposed method}
& Valid regression inference under Cox process models
& Working Poisson model on discretized counts; latent dependence handled in asymptotic theory
& No parametric pair-correlation specification
& Yes (penalization + de-biasing under increasing-domain-type asymptotics) \\

\bottomrule
\end{tabular}

\caption{Comparison of inference strategies for Cox and related spatial point process models. Methods differ in inferential target, treatment of spatial dependence, need for second-order modeling specification, and support for high-dimensional inference.}
\label{tab:lit-comparison}

\end{table}

\end{landscape}

\section{Additional Numerical Results}
\label{app:addn_rslts}

\subsection{{Expanded Simulations}}
{Figure~\ref{fig:sim-ci-width} visualizes the width of confidence intervals from different inference procedures in the simulation. As expected, penalized PMLE has relatively wider confidence intervals than Scampr and the Bayesian models, but in general on a comparable magnitude.}

\begin{figure}
    \centering
    \includegraphics[width=17cm]{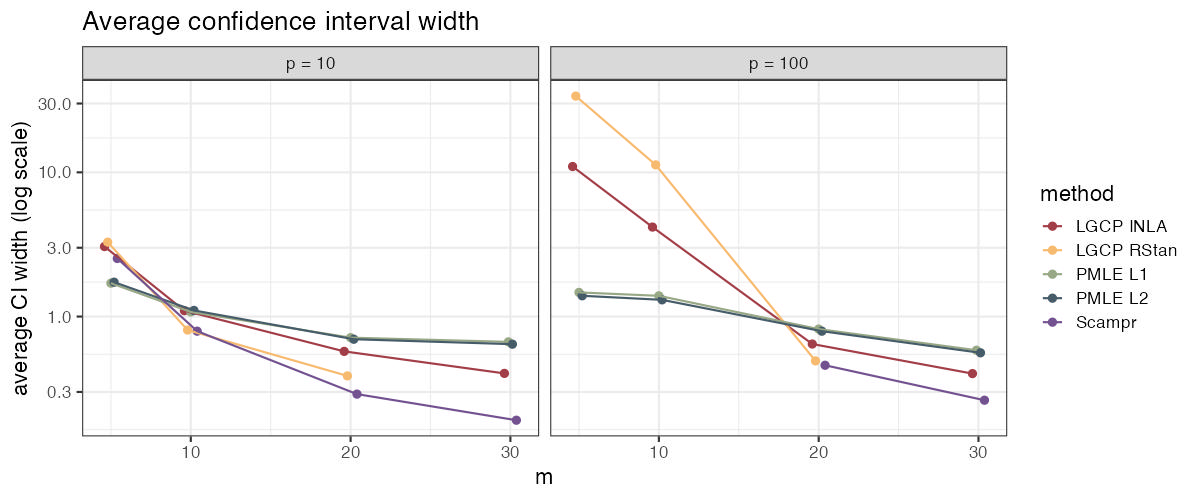}
    \caption{{Average confidence interval width from each method in the simulation}}
    \label{fig:sim-ci-width}
\end{figure}

{In addition to the non-stationary and heteroskedastic simulation setting studied in Section~\ref{sec:sim}, we also compare the same set of models under a vanilla LGCP data generating mechanism. In this study, the data is generated from a similar process as Section~\ref{sec:sim}, except that the Gaussian random field is stationary with an exponential covariance structure. Figure~\ref{fig:sim-vanilla-dgm} summarizes each model's inference performance under this setting. With a correctly specified model, INLA and PMLE both achieve the nominal $95\%$ coverage. Specifically, INLA demonstrates higher power whereas PMLE was more conservative and approaches the nominal level relatively slower, due to its over-parametrized and semi-parametric nature. However, we did not observe a significant gap in power or precision despite the conservativeness of PMLE.}

\begin{figure}[ht]
    \centering
    \includegraphics[width=17cm]{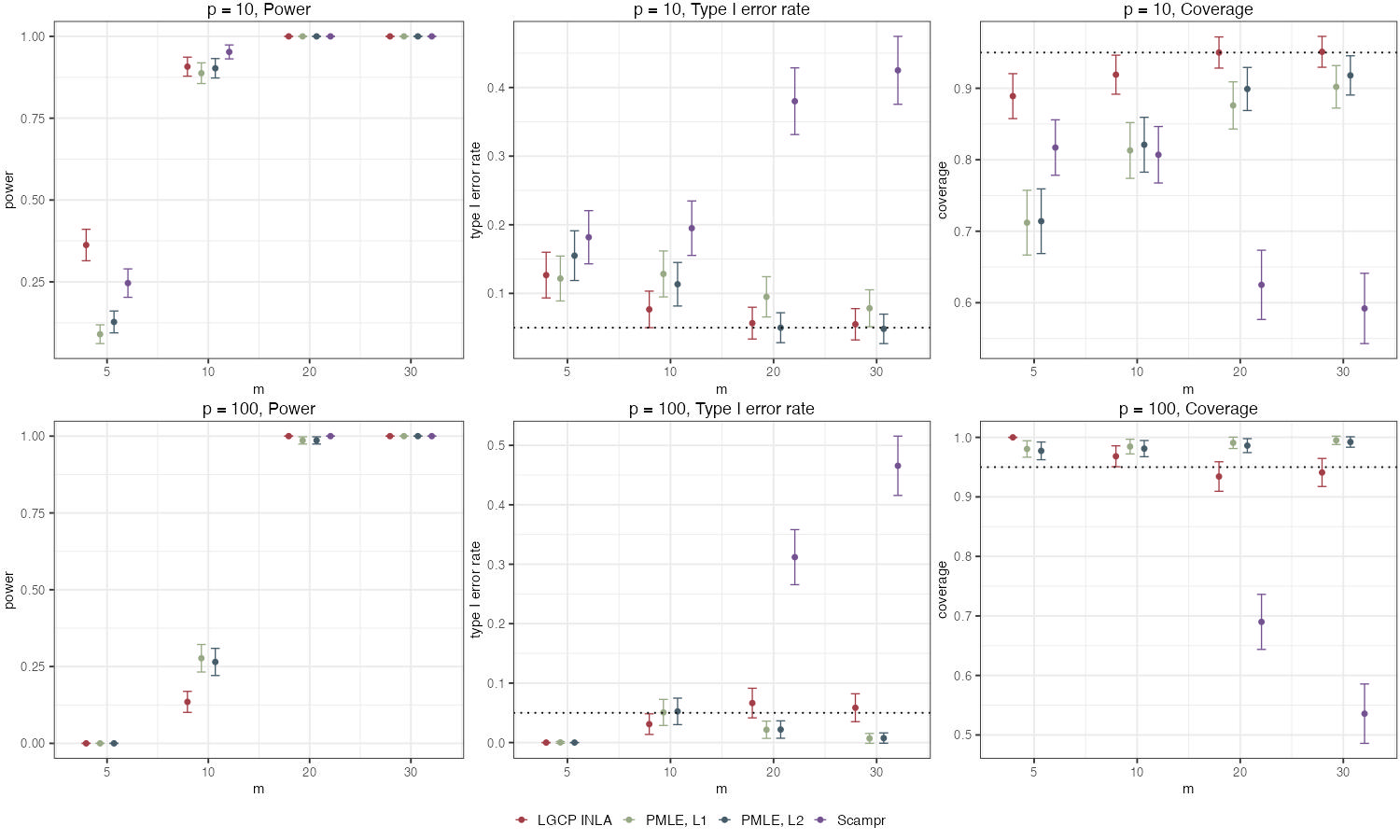}
    \caption{{Comparison of coverage, type I error rate and power for penalized PMLE, Scampr and Bayesian LGCP methods, under the vanilla LGCP data generating mechanism.}}
    \label{fig:sim-vanilla-dgm}
\end{figure}

{All models achieved comparable estimation accuracy, as reflected in Table~\ref{tab:sim-expanded-mse}. Hence, the inflated type I error rate and lower coverage of Scampr was not due to inaccurate estimation, but instead underestimated variability---similar to the main simulation study in Section~\ref{sec:sim}. This can be seen, for example, from Scampr’s median CI width of 0.224 in the $p=100, m=30$ case, in comparison to 0.430 for INLA.}

\begin{table}[]
\centering
\begin{tabular}{@{}lllll@{}}
\toprule
        & PMLE L1 & PMLE L2 & INLA   & Scampr \\ \midrule
$p=10$  & 0.0126  & 0.0103  & 0.0105 & 0.0164 \\
$p=100$ & 0.0122  & 0.0107  & 0.0128 & 0.0230 \\ \bottomrule
\end{tabular}
\caption{{Average entry-wise estimation MSE under the $n=30^2$ scenario for all models, under the vanilla LGCP data generating mechanism (LGCP with \texttt{RStan} dropped due to intensive computation)}}
\label{tab:sim-expanded-mse}
\end{table}

We also report supplementary simulation results, further providing insights on the mechanism of regularization.
Figure~\ref{fig:sim-cv-surface} presents the cross-validation surface for the tuning parameters $(\gamma,\tau)$ under the three graph specifications considered in the sensitivity analysis. Across these specifications, the cross-validation surface is well-conditioned, with a clear joint optimum in the interior region rather than noisy or ridge-like tradeoffs between the two tuning parameters. Moreover, the MSE remains reasonably similar in the immediate neighborhood around the selected optimum, indicating that the cross-validation procedure is stable under these alternative graph structures.

\begin{figure}
    \centering
    \includegraphics[width=17cm]{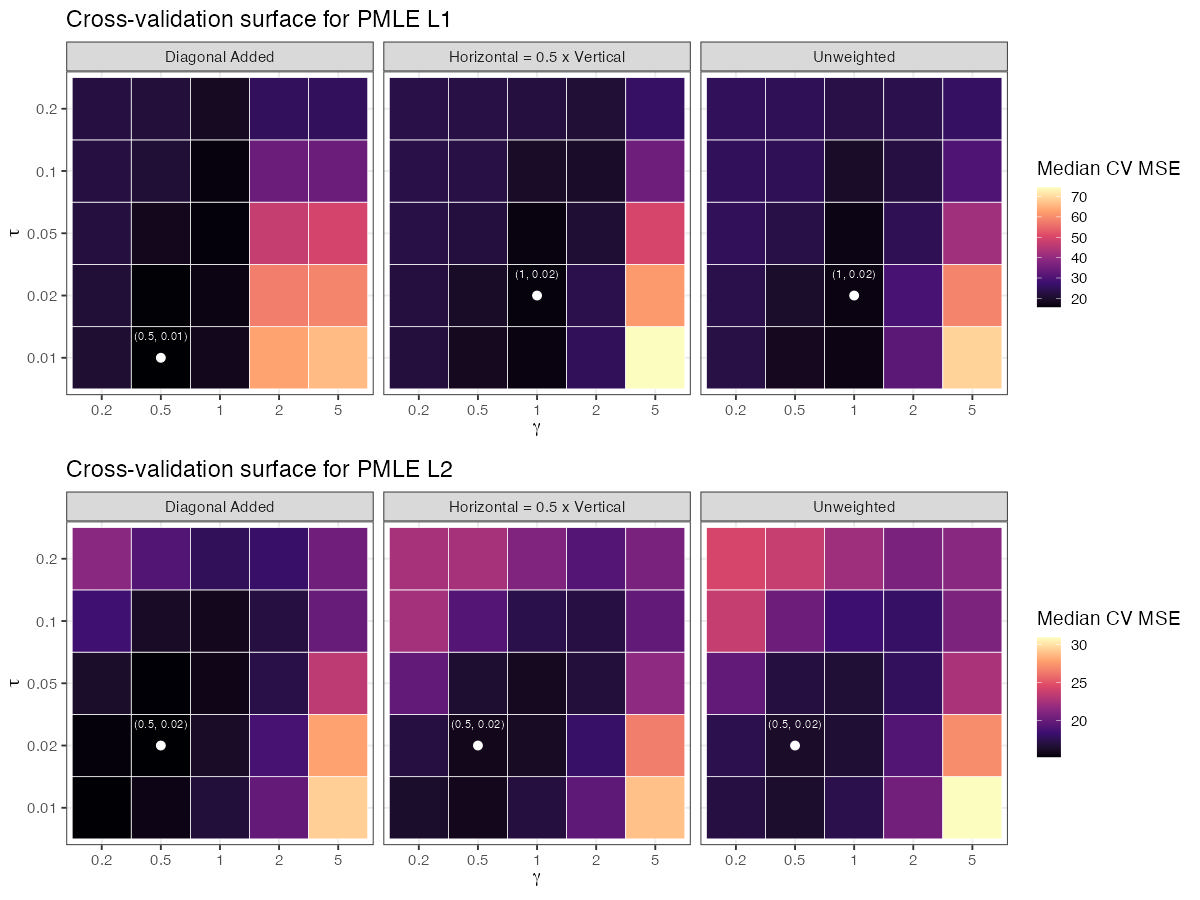}
    \caption{Cross-validation surface for $(\gamma,\tau)$ in the $p=100, m = 10^2$ scenario under different graph specifications.}
    \label{fig:sim-cv-surface}
\end{figure}

Figures~\ref{fig:sim-base-int-gap} and~\ref{fig:sim-sparsity} present additional numerical analyses beyond graph specification. 
Figure~\ref{fig:sim-base-int-gap} examines the interpretability of the baseline intensity estimation through the convergence trajectory under our specified asymptotic regime. Figure~\ref{fig:sim-sparsity} illustrates the degree-of-freedom control behavior under $\ell_1$ and $\ell_2$ fusion penalties.

{Since the incidents of the point process are generated from a fine grid, the baseline intensities in each grid serve as an approximate ``ground truth'' to compare the fitted baseline intensity values. Here we visualize the root mean squared error (RMSE) between the fitted and true intensity parameters. Note that we expect the fitted baseline intensities to be biased due to the target parameter absorbing the error term induced by double stochasticity (recall the formulation in Lemma~\ref{lem:target-param}); nevertheless, the convergence trajectory will provide interpretability on the model's behavior under our specified asymptotic regime underlying the spatial domain. Figure~\ref{fig:sim-base-int-gap} shows the converging trends on the baseline intensity estimation under the specified asymptotic regime, and further varies the existence of the bias term as the RMSE trends converge to a non-zero level.}

\begin{figure}
    \centering
    \includegraphics[width=17cm]{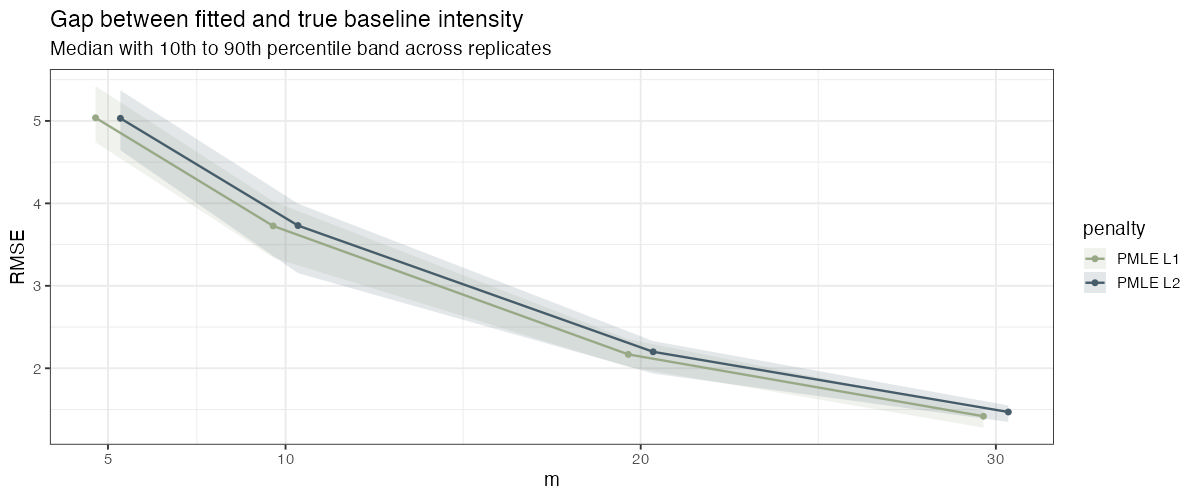}
    \caption{{RMSE between the fitted and true baseline intensity parameters with the optimal tuning parameters, under the $p=100$ scenario with increasing sample sizes. }}
    \label{fig:sim-base-int-gap}
\end{figure}

{Figure~\ref{fig:sim-sparsity} further demonstrates how the $\ell_1$ and $\ell_2$ fusion penalties induce ``smoothness'' in the estimated intensity surface. Here we consider two notions of smoothness: }
\begin{itemize}
    \item Active edge fraction, defined as the fraction of edges whose connecting nodes have different (using $10^{-3}$ as a numerical threshold) estimated baseline intensity values, that is,
    \begin{equation*}
        \frac{1}{|E|}\sum_{j=1}^{|E|} \mathbbm{1}\left\{ B_{j\cdot} \hat{\balpha} > 10^{-3} \right\};
    \end{equation*}
    \item Normalized edge-difference $\ell_2$ norm, which measures the overall magnitudes by which the baseline estimates between connected nodes differ; more formally
    \begin{equation*}
        \left(\frac{1}{|E|}\sum_{j=1}^{|E|} (B_{j\cdot}\hat{\balpha})^2 \right)^{1/2}
    \end{equation*}
\end{itemize}
where we recall $E$ indicates the edge set, and $B$ denotes the edge incidence matrix of the graph.
{Active edge fraction captures exact piecewise-constant structure in the estimated baseline intensity surface, and the normalized edge-difference $\ell_2$ norm captures ``soft'' smoothness of the surface. $\ell_1$ fusion is expected to drive both smoothness measures to 0, whereas $\ell_2$ fusion only effectively reduces the edge-difference $\ell_2$ norm, as the fusion penalty driven by $\gamma$ becomes stronger. Figure~\ref{fig:sim-sparsity} illustrates that both the $\ell_1$ and $\ell_2$ fusion penalty of PMLE controls the degree-of-freedom of the intensity surface in the expected way across different sample sizes under the asymptotic regime.}

\begin{figure}[ht]
    \centering
    \includegraphics[width=17cm]{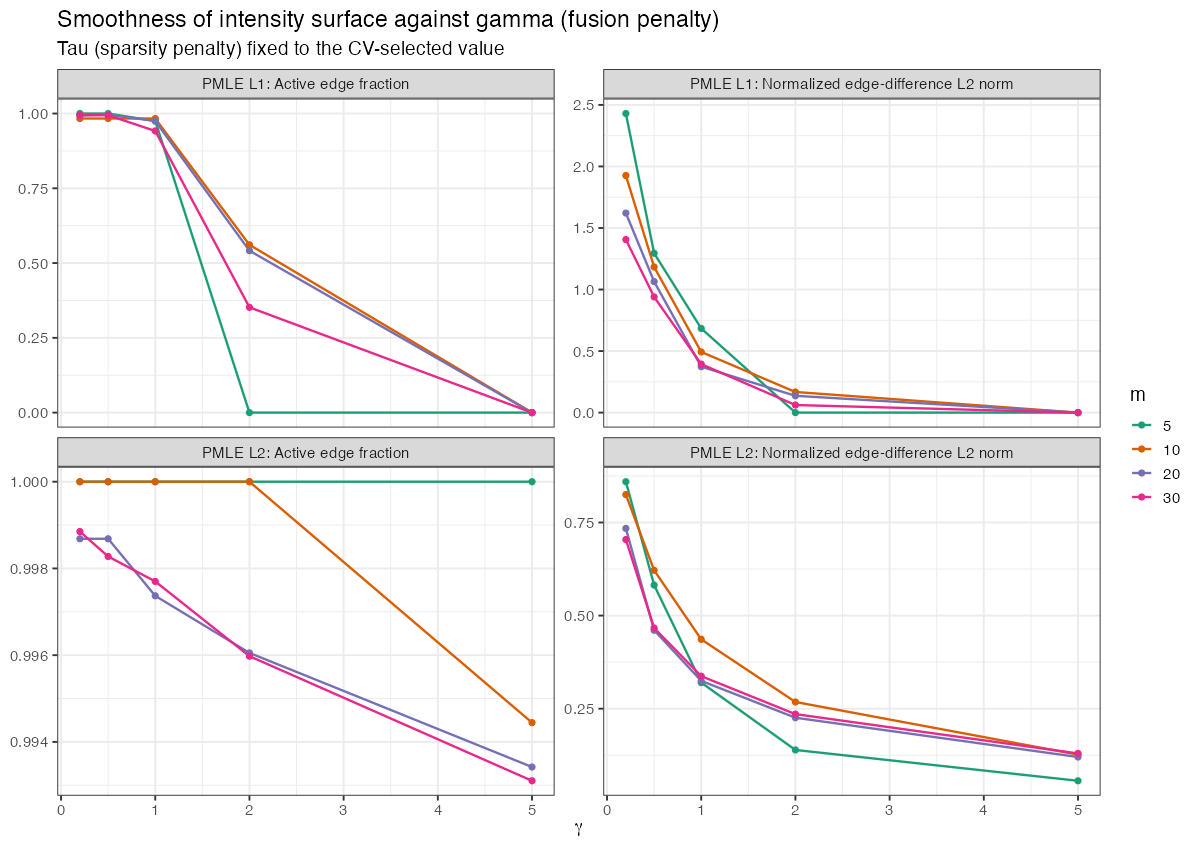}
    \caption{{Smoothness of the fitted baseline intensity surface against the value of fusion penalty $\gamma$ under different sample sizes $m$, with sparsity penalty $\tau$ fixed to the optimal value in cross-validation.}}
    \label{fig:sim-sparsity}
\end{figure}

\subsection{Spatial Confounding Analysis with Seattle Crime Data}
\label{app:spatial-confounding}

A common concern in the analysis of spatial data is the effect of
\emph{spatial confounding} \citep{reich2006effects, paciorek2010importance}.
The presence of spatial confounding, which occurs when covariates contributing to
the variability in the response are spatially structured, may introduce biases to
the estimated effect sizes. To investigate the sensitivity of each inference
procedure to spatial confounding as well as their performance with high-dimensional
covariates, we fit each model with 50 simulated noise variables included alongside
the original covariates. Among the 50 noise variables, the first 10 are drawn from
graph-structured Gaussian random fields and hence spatially structured, and the
remaining 40 are drawn from independent standard Normal distributions.

Figure~\ref{fig:data-par-diff} compares the estimated effects along with 95\% CI/CrI
before and after the high-dimensional synthetic noise variables are introduced. We
observe that PMLE under both $\ell_1$ and $\ell_2$ fusion has robust performance
when covariate dimensionality increases and when spatially structured noise is
introduced. In contrast, the Bayesian methods and Scampr all show some level of
sensitivity, reflected by widening CrIs, for example for medium basin and fire
station under both BYM2 and LGCP, or by marginally significant sign changes, for
example for police station under Scampr, when high-dimensional noise variables are
introduced.

Another assessment of model robustness can be learned from the estimated effects
for the noise variables, where a robust model is expected to correctly identify
them as having weak effects. Figure~\ref{fig:data-noise-est} shows this comparison,
where PMLE estimates near-zero effects for all noise variables with tight CIs. All
Bayesian methods and Scampr falsely identify several noise variables as
statistically significant, likely due to lack of explicit regularization on both the
covariate effects and the baseline intensity surface.

\begin{figure}
    \centering
    \includegraphics[width=17cm]{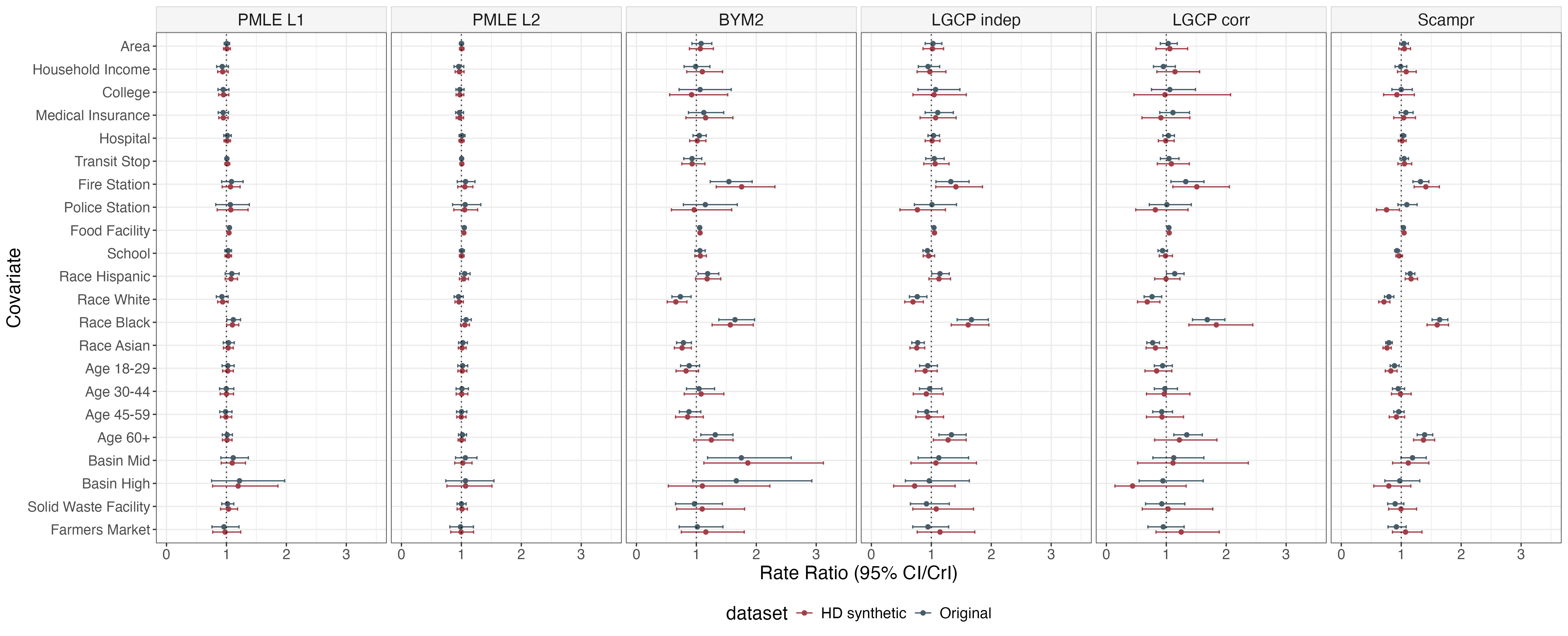}
    \caption{Estimated effects for the original set of covariates, before and after the high-dimensional synthetic noise variables are introduced.}
    \label{fig:data-par-diff}
\end{figure}

\begin{figure}
    \centering
    \includegraphics[width=17cm]{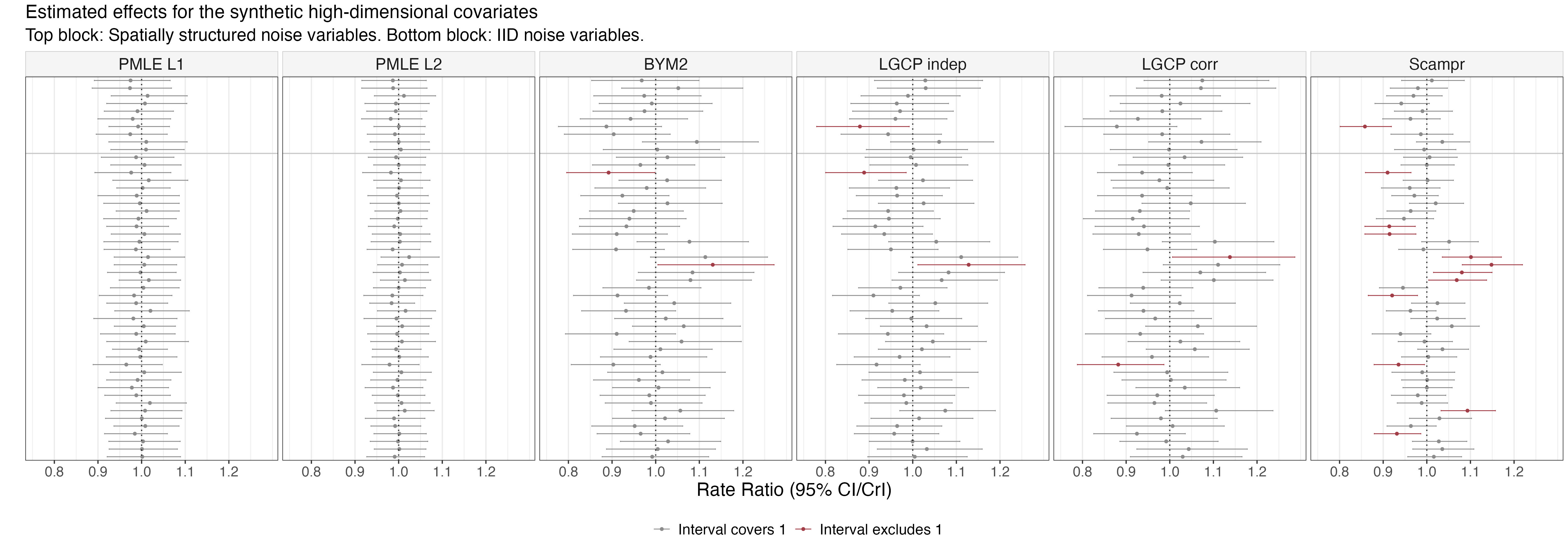}
    \caption{Estimated effects for the high-dimensional synthetic noise variables. The top block represents the estimated effects of spatially structured noise variables, and the bottom block corresponds to the independent noise variables.}
    \label{fig:data-noise-est}
\end{figure}

\section{Technical Details and Proofs}
\label{app:proofs}

This section includes proofs for our theoretical claims in Sections~\ref{sec:model} and~\ref{sec:inference}. 
We reintroduce our notation for clarity. The true, continuous baseline intensity is denoted as $\alpha^0(\cdot)$, and the true regression parameters are denoted as $\boldsymbol\beta^0$. We denote the discretized baseline vector, i.e. $\alpha^0(\cdot)$ evaluated at locations $\boldsymbol s=(s_1, \ldots, s_n)$, as ${\boldsymbol{\tilde\alpha}(\boldsymbol s)}$ to distinguish it from the baseline intensity function. We also define $\phi(s)=\log\mathbb E_0[\exp\varepsilon(s)]$, $\boldsymbol B = (\lvert\Omega_1\rvert, \ldots, \lvert\Omega_n\rvert)$, 
and recall that $\ell(\cdot)$ is the Possion log-likelihood as defined in Section~\ref{sec:model}.

Empirical process notations are adopted, where under discretization of the observation window $\Omega$, we denote $\mathbb P_0 f(\theta;X,Y):= \mathbb E_0[f(\theta;X, Y)]$ with $\mathbb E_0$ being the expectation taken under the true distribution of $X, Y$, 
and $\mathbb P_nf(\theta;X,Y):= n^{-1}\sum_i f(\theta;X_i, Y_i)$.

We first prove Lemma~\ref{lem:target-param} by examining the relationship between the target parameter $\boldsymbol\theta^\dagger:=(\boldsymbol{\tilde\alpha}^\dagger, \boldsymbol\beta^\dagger)$, which is the solution to
\[
    -\nabla_{(\boldsymbol{\tilde\alpha}, \boldsymbol\beta)} \mathbb P_0\ell(\boldsymbol{\tilde\alpha}, \boldsymbol\beta) 
    = 0,
\]
and the true parameter $\boldsymbol\beta^0$ along with the function $\alpha^0(\cdot)$ underlying the Cox process. In particular, we show that the Poisson likelihood yields an unbiased estimating equation for $\boldsymbol\beta$ despite the ignored error random field as well as misspecification of $\alpha^0(\cdot)$. With the fusion penalty $R(\boldsymbol{\tilde\alpha};\mathcal G_n)$ incorporated into the objective function, we further bound the gap between the penalized solution $\boldsymbol\theta^*$ and $\boldsymbol\theta^\dagger$ under different conditions on the smoothness of $\alpha^0(\cdot), \phi(\cdot)$ and $X(\cdot)$.

We then use empirical process arguments to show the convergence of the penalized PMLE to the target parameters, following a similar outline as in \citet{haris2019generalized} and \citet{sondhi2024doubly}, with an adaptation to the heavy-tailed distribution of the observations in our setting due to double stochasticity.

Finally, we establish the asymptotic linearity of the de-biased estimator $\hat{\boldsymbol b}$ and, in turn, show the validity of our variance estimator along with the inference procedure. 

\subsection{Consistency}

\begin{proof}[Proof of Lemma~\ref{lem:target-param}]

For region $\Omega_i$, the working Poisson log-likelihood satisfies
\[
    \frac{\partial \ell}{\partial \tilde\alpha_i}
    =
    Y_i
    -
    |\Omega_i|
    \exp\left\{
        \tilde\alpha_i
        +
        X_i^\top\boldsymbol\beta
    \right\}.
\]
Hence by Fubini's theorem under Assumption~\ref{assump:regularity},
\begin{align}
    \notag -\mathbb P_0
    \frac{\partial \ell}{\partial \tilde\alpha_i}
    &=
    -
    \mathbb E_0[Y_i]
    +
    |\Omega_i|
    \exp\left\{
        \tilde\alpha_i
        +
        X_i^\top\boldsymbol\beta
    \right\} \\
    &= -\int_{\Omega_i}
    \exp\left\{
        \alpha^0(s)
        +
        X(s)^\top\boldsymbol\beta^0
        +
        \phi(s)
    \right\} \dd s
    + |\Omega_i|
    \exp\left\{
        \tilde\alpha_i
        +
        X_i^\top\boldsymbol\beta
    \right\},
    \label{eq:true-cell-mean-absorbed}
\end{align}
where $ \phi(s)
    =
    \log
    \mathbb E_0
    \left[
        \exp\{\varepsilon(s)\}
    \right]$. Define the projected nuisance target as
\[
    \alpha_i^\ddagger
    :=
    \log\left[
        |\Omega_i|^{-1}
        \int_{\Omega_i}
        \exp\left\{
            \alpha^0(s)
            +
            \phi(s)
            +
            \left(
                X(s)-X_i
            \right)^\top
            \boldsymbol\beta^0
        \right\}
        \dd s
    \right].
    \label{eq:alpha-dagger-projected}
\]
Then, the mean value theorem for integrals together with Assumption~\ref{assump:regularity} imply the existence of  $s_i^*\in\Omega_i$ for all $i = 1, 2, \ldots,n$ such that 
\[
    \alpha_i^\dagger(s_i^*) = \alpha_i^\ddagger.
\]
Hence, we have
\begin{align}
    |\Omega_i|
    \exp\left\{
        \alpha_i^\dagger(s_i^*)
        +
        X_i^\top\boldsymbol\beta^0
    \right\}
    &=
    \int_{\Omega_i}
    \exp\left\{
        \alpha^0(s)
        +
        X(s)^\top\boldsymbol\beta^0 +
        \phi(s)
    \right\}
    \dd s = \mathbb E_0[Y_i].
    \label{eq:projected-target-mean-match}
\end{align}
Combining
\eqref{eq:true-cell-mean-absorbed} and
\eqref{eq:projected-target-mean-match}, we obtain
\begin{align}
    -
    \mathbb P_0
    \frac{\partial \ell}{\partial \tilde\alpha_i}
    &=
    -
    |\Omega_i|
    \exp\left\{
        \alpha_i^\dagger(s_i^*)
        +
        X_i^\top\boldsymbol\beta^0
    \right\}
    +
    |\Omega_i|
    \exp\left\{
        \tilde\alpha_i
        +
        X_i^\top\boldsymbol\beta
    \right\}.
    \label{eq:alpha-score-projected}
\end{align}
Therefore,
\[
    -
    \frac{\partial}{\partial \tilde\alpha_i}
    \mathbb P_0
    \ell(
        \tilde{\boldsymbol\alpha},
        \boldsymbol\beta
    )
    \bigg|_{
        (
            \boldsymbol\alpha^\dagger,
            \boldsymbol\beta^0
        )
    }
    =
    0.
\]
Moreover, because the working likelihood uses the region-level covariates
$X_i$, its population score with respect to $\boldsymbol\beta$ satisfies
\[
    -
    \nabla_{\boldsymbol\beta}
    \mathbb P_0
    \ell(
        \tilde{\boldsymbol\alpha},
        \boldsymbol\beta
    )
    =
    -
    \sum_{i=1}^n
    X_i
    \frac{\partial}{\partial \tilde\alpha_i}
    \mathbb P_0
    \ell(
        \tilde{\boldsymbol\alpha},
        \boldsymbol\beta
    ).
\]
Consequently,
\[
    -
    \nabla_{\boldsymbol\beta}
    \mathbb P_0
    \ell(
        \tilde{\boldsymbol\alpha},
        \boldsymbol\beta
    )
    \bigg|_{
        (
            \boldsymbol\alpha^\dagger,
            \boldsymbol\beta^0
        )
    }
    =
    0.
\]
Together, the preceding displays establish
\[
    -
    \nabla_{
        (
            \tilde{\boldsymbol\alpha},
            \boldsymbol\beta
        )
    }
    \mathbb P_0
    \ell(
        \tilde{\boldsymbol\alpha},
        \boldsymbol\beta
    )
    \bigg|_{
        (
            \boldsymbol\alpha^\dagger,
            \boldsymbol\beta^0
        )
    }
    = -
    \nabla_{
        (
            \tilde{\boldsymbol\alpha},
            \boldsymbol\beta
        )
    }
    \mathbb P_0
    \ell(
        \tilde{\boldsymbol\alpha},
        \boldsymbol\beta
    )
    \bigg|_{
        (
            \boldsymbol\alpha^\dagger,
            \boldsymbol\beta^\dagger
        )
    } =   0.
\]
\end{proof}

Throughout this section, we denote the smooth portion of our objective function as
\[
    \mathcal L(\boldsymbol\theta) := -\ell(\boldsymbol{\tilde\alpha},\boldsymbol\beta) + \gamma_n R(\boldsymbol{\tilde\alpha};\mathcal G_n).
\]

\begin{proof}[Proof of Lemma~\ref{lem:target-param-fusion}]
Recall the definition of the linear predictor $ \boldsymbol\eta
    := (\eta_1,\ldots, \eta_n)
    = \btalpha+\bX\bbeta.$
For the working Poisson negative log-likelihood, up to constants not
depending on \((\btalpha,\bbeta)\),
\[
    -\mathbb P_0\ell(\btalpha,\bbeta)
    \propto
    \sum_{i=1}^n
    \left[
        |\Omega_i|
        \exp(\eta_i)
        -
        \mathbb E_0(Y_i)\eta_i
    \right].
\]
Thus, $-\mathbb P_0\ell(\btalpha,\bbeta)$ depends on \((\btalpha,\bbeta)\) only through
\(\boldsymbol\eta=\btalpha+\bX\bbeta\). With a slight abuse of notation, we now write $-\mathbb P_0\ell(\boldsymbol\eta)$ as a function of $\boldsymbol{\eta}$. By the optimality of $\btheta^\dagger$ with respect to $-\mathbb P_0\ell(\cdot)$ established in Lemma~\ref{lem:target-param}, we have
\[
    -\nabla_{\boldsymbol\eta}
    \mathbb P_0\ell(\boldsymbol\eta^\dagger)
    =
    0,
    \text{ where }
    \boldsymbol\eta^\dagger
    =
    \tilde{\boldsymbol\alpha}^{\dagger}
    +
    \bX\bbeta^0.
\]
Let
$
    \boldsymbol\eta^*
    :=
    {\boldsymbol\alpha}^*
    +
    \bX\bbeta^*
$ be the linear predictor associated with $(\balpha^*, \bbeta^*)$. Due to its optimality for $\mathbb P_0\mathcal L(\btheta)$ with respect to $\tilde{\balpha}$, and since $\partial\boldsymbol{\eta}/\partial{\tilde\balpha} = I_n$, we have
\begin{equation}
    -\nabla_{\boldsymbol\eta}\mathbb P_0\ell(\boldsymbol\eta^*)
    +
    \gamma_n\nabla_{\btalpha}R(\tilde{\boldsymbol\alpha}^*;\mathcal G_n)
    =
    0.
    \label{equ:foc-alpha}
\end{equation}
And evaluating the first-order condition with respect to $\bbeta$ yields
\begin{equation}
    \bX^\top
    \nabla_{\boldsymbol\eta}\mathbb P_0\ell(\boldsymbol\eta^*)
    =
    0.
    \label{equ:foc-beta}
\end{equation}
Multiplying both sides of \eqref{equ:foc-alpha} on the left by \(\bX^\top\) and comparing with \eqref{equ:foc-beta} yields
\begin{equation}
    \bX^\top
    \nabla_{\btalpha}R(\tilde{\boldsymbol\alpha}^*;\mathcal G_n) = 0.
    \label{equ:fusion-grad}
\end{equation}

Now define
\[
    \Delta_\beta
    :=
    \bbeta^*-\bbeta^0,
    \qquad
    \Delta_\eta
    :=
    \boldsymbol\eta^*-\boldsymbol\eta^\dagger;
\]
it then follows that
\begin{equation}
    {\boldsymbol\alpha}^*
    -
    {\boldsymbol\alpha}^{\dagger}
    =
    \Delta_\eta-\bX\Delta_\beta.
    \label{equ:alpha-gap-wrt-beta}
\end{equation}
By the definition of $\bar H_n(\tilde{\boldsymbol\alpha})$ in Assumption~\ref{assump:graph-curv},
\[
    \gamma_n\nabla_{\tilde\balpha} R({\boldsymbol\alpha}^*;\mathcal G_n)
    -
    \gamma_n\nabla_{\tilde\balpha} R({\boldsymbol\alpha}^\dagger;\mathcal G_n)
    =
    \bar H_n({\boldsymbol\alpha}^*)
    \left(
        {\boldsymbol\alpha}^*
        -
        {\boldsymbol\alpha}^{\dagger}
    \right).
\]
Therefore,
\[
     \gamma_n\nabla_{\tilde\balpha} R({\boldsymbol\alpha}^*;\mathcal G_n)
    =
    G_n^\dagger
    +
    \bar H_n({\boldsymbol\alpha}^*)
    \left(
        \Delta_\eta-\bX\Delta_\beta
    \right).
\]
Substituting this expression into \eqref{equ:fusion-grad} leads to
\begin{equation}
    \frac{1}{n}
    \bX^\top
    \bar H_n({\boldsymbol\alpha}^*)
    \bX
    \Delta_\beta
    =
    \frac{1}{n}
    \bX^\top
    G_n^\dagger
    +
    \frac{1}{n}
    \bX^\top
    \bar H_n({\boldsymbol\alpha}^*)
    \Delta_\eta:= W_n.
    \label{equ:delta-equ}
\end{equation}
We write the right-hand-side (RHS) of \eqref{equ:delta-equ} as $W_n$ for simplicity.
Under Assumptions~\ref{assump:pen-scale}-\romannumeral3) and~\ref{assump:graph-curv}-\romannumeral1), it holds that $\lVert W_n\rVert_\infty = O_P(\sqrt{n^{-1}\log p})$.

Further, multiplying both sides of \eqref{equ:delta-equ} on the left by \(\Delta_\beta^\top\), we obtain
\begin{equation}
    \frac{1}{n}
    \Delta_\beta^\top
    \bX^\top
    \bar H_n({\boldsymbol\alpha}^*)
    \bX
    \Delta_\beta
    =
    \Delta_\beta^\top W_n
    \le \|\Delta_\beta\|_1
    \|W_n\|_\infty,
    \label{equ:delta-beta-norm}
\end{equation}
where the `$\le$' follows from Holder's inequality.

We next examine the left-hand-side (LHS) of \eqref{equ:delta-beta-norm} aiming to bound $\Delta_\beta$ under $\ell_1$ and $\ell_2$ norms respectively.

\textbf{$\ell_1$-norm of $\Delta_\beta$:} Following from the $\ell_1$ fusion compatibility condition of Assumption~\ref{assump:graph-curv}-\romannumeral2),
\[
    \frac{\kappa_{F}}{q}
    \|\Delta_\beta\|_1^2
    \le
    \frac{1}{n}
    \Delta_\beta^\top
    \bX^\top
    \bar H_n({\boldsymbol\alpha}^*)
    \bX
    \Delta_\beta
    \le
    \|\Delta_\beta\|_1
    \|W_n\|_\infty,
\]
which leads to
\[
    \|\bbeta^*-\bbeta^0\|_1 = \|\Delta_\beta\|_1
    \le
    \frac{q}{\kappa_{F}}
    \|W_n\|_\infty
    = O_P\left(q\sqrt{\frac{\log p}{n}}\right).
\]

\textbf{$\ell_2$-norm of $\Delta_\beta$:}
Likewise, using the \(\ell_2\) compatibility condition in
Assumption~\ref{assump:graph-curv}-\romannumeral2), we have
\[
    \kappa_{F}
    \|\Delta_\beta\|_2^2
    \le
    \frac{1}{n}
    \Delta_\beta^\top
    \bX^\top
    \bar H_n({\boldsymbol\alpha}^*)
    \bX
    \Delta_\beta
     \le \|\Delta_\beta\|_1
    \|W_n\|_\infty,
\]
which, combined with the established $\ell_1$ bound of $\Delta_\beta$ above, yields
\[
    \|\bbeta^*-\bbeta^0\|_2 =
    \|\Delta_\beta\|_2 =
    O_P\left(
        \sqrt q\cdot \sqrt{\frac{\log p}{n}}
    \right).
\]

We next analyze the gap $\Delta_\alpha$. Recall from \eqref{equ:alpha-gap-wrt-beta} that
\[
    \Delta_\alpha
    =
    \Delta_\eta-\bX\Delta_\beta.
\]
We continue from \eqref{equ:foc-alpha} by adding $\nabla_\eta\mathbb P_0
    \ell(\boldsymbol\eta^\dagger)$, which is zero based on Lemma~\ref{lem:target-param}:
\begin{equation}
    -
    \nabla_\eta\mathbb P_0
    \ell(\boldsymbol\eta^*)
    +
    \nabla_\eta\mathbb P_0
    \ell(\boldsymbol\eta^\dagger)
    +
    \gamma_n\nabla R(\boldsymbol\alpha^*;\mathcal G_n)
    =
    0.
    \label{equ:score-cond}
\end{equation}
For the working Poisson likelihood,
\(
    \nabla_\eta^2
    \{
        -\mathbb P_0\ell(\boldsymbol\eta)
    \}
    =
    \operatorname{diag}
    \left\{
        |\Omega_i|\exp(\eta_i)
    \right\}_{i=1}^n.
\)
Then for the $i$th component,
\begin{align*}
    \left[
        -\nabla_\eta\mathbb P_0
        \ell(\boldsymbol\eta^*)
        +
        \nabla_\eta\mathbb P_0
        \ell(\boldsymbol\eta^\dagger)
    \right]_i
    & =
    |\Omega_i|
    \left\{
        \exp(\eta_i^*)-\exp(\eta_i^\dagger)
    \right\}
    \\
    &    =
    |\Omega_i|
    (\eta_i^*-\eta_i^\dagger)
    \int_0^1
    \exp\left[
        \eta_i^\dagger
        +
        t(\eta_i^*-\eta_i^\dagger)
    \right]\dd t
\end{align*}
by the mean value theorem. Define $\bar W_n
    :=
    \operatorname{diag}(\bar w_1,\ldots,\bar w_n)$, where
\[
    \bar w_i
    :=
    |\Omega_i|
    \int_0^1
    \exp\left[
        \eta_i^\dagger
        +
        t(\eta_i^*-\eta_i^\dagger)
    \right]\dd t,
    \qquad
    i=1,\ldots,n.
\]
Then, we have
\begin{equation}
    -
    \nabla_\eta\mathbb P_0
    \ell(\boldsymbol\eta^*)
    +
    \nabla_\eta\mathbb P_0
    \ell(\boldsymbol\eta^\dagger)
    =
    \bar W_n
    (
        \boldsymbol\eta^*
        -
        \boldsymbol\eta^\dagger
    ).
    \label{equ:mvt-eta}
\end{equation}

Also, recall that by the definition of generalized Hessian in Assumption~\ref{assump:graph-curv},
\begin{equation}
    \gamma_n\nabla R(\boldsymbol\alpha^*;\mathcal G_n)
    =
    G_n^\dagger
    +
    \bar H_n(\boldsymbol\alpha^*)\Delta_\alpha;
    \label{equ:gen-hess}
\end{equation}
hereinafter, we write $\bar H_n := \bar H_n(\boldsymbol\alpha^*)$ for simplicity. 
Combining \eqref{equ:score-cond}, \eqref{equ:mvt-eta} and \eqref{equ:gen-hess} yields
\[
    \bar W_n\Delta_\eta
    +
    G_n^\dagger
    +
    \bar H_n
    (
        \Delta_\eta-\bX\Delta_\beta
    ) =
    \bar W_n\Delta_\eta
    +
    G_n^\dagger
    +
    \bar H_n\Delta_\alpha
    = 
    0,
\]
and hence
\begin{equation}
    (
        \bar W_n+\bar H_n
    )
    \Delta_\eta
    =
    -
    G_n^\dagger
    +
    \bar H_n\bX\Delta_\beta.
    \label{A.5}
\end{equation}

By Assumption~\ref{assump:bounded-cont-intensity},
\(
    \lambda_{\min}(\bar W_n)
    \ge
    \psi.
\)
And further since \(\bar H_n\) is symmetric positive semidefinite under Assumption~\ref{assump:graph-curv}-\romannumeral3), we have $\lambda_{\min}(\bar W_n+\bar H_n)
    \ge
    m_\eta$
and $|
        (
            \bar W_n+\bar H_n
        )^{-1}
    \|_{\operatorname{op}}
    \le
    \psi^{-1}$.
Taking $\ell_2$ norms in \eqref{A.5}:
\[
    \|\Delta_\eta\|_2
    \le
    \psi^{-1}
    \left[
        \|G_n^\dagger\|_2
        +
        \|
            \bar H_n\bX\Delta_\beta
        \|_2
    \right],
\]
where, by Assumption~\ref{assump:graph-curv}-\romannumeral3),
\[
    \|
        \bar H_n\bX\Delta_\beta
    \|_2
    \le
    \|
        \bar H_n\bX
    \|_{\operatorname{op}}
    \|\Delta_\beta\|_2.
\]
Thus, the $\ell_2$ norm of $\Delta_\eta$ follows
\begin{equation}
    \frac{1}{\sqrt n}
    \|\Delta_\eta\|_2
    =
    O_P\left(
        \rho_{2,n}
        +
        \kappa_{2,n}
        \sqrt{\frac{q\log p}{n}}
    \right).
    \label{A.6}
\end{equation}
Recalling \eqref{equ:alpha-gap-wrt-beta} again, triangular inequality yields
\[
    \frac{1}{\sqrt n}
    \|\Delta_\alpha\|_2
    \le
    \frac{1}{\sqrt n}
    \|\Delta_\eta\|_2
    +
    \frac{1}{\sqrt n}
    \|\bX\Delta_\beta\|_2;
\]
and furthermore, under Assumption~\ref{assump:graph-curv},
\begin{equation}
    \frac{1}{\sqrt n}
    \|\bX\Delta_\beta\|_2
    \le
    \frac{1}{\sqrt n}
    \|\bX\|_{\operatorname{op}}
    \|\Delta_\beta\|_2
    =
    O_P\left(
        \sqrt{\frac{q\log p}{n}}
    \right).
    \label{A.7}
\end{equation}
Combining \eqref{A.6} and \eqref{A.7}, we obtain
\[
    \frac{1}{\sqrt n}
    \|
        \boldsymbol\alpha^*-\boldsymbol\alpha^\dagger
    \|_2
    =
    O_P\left(
        \rho_{2,n}
        +
        (1+\kappa_{2,n})
        \sqrt{\frac{q\log p}{n}}
    \right).
\]

Next, we prove the $\ell_1$ bound of $\Delta_\alpha$ when the last part of Assumption~\ref{assump:graph-curv}-\romannumeral3) holds. 
Taking $\ell_1$ norm in \eqref{A.5} yields
\[
    \|\Delta_\eta\|_1
    =
    O_P\left(
        \|G_n^\dagger\|_1
        +
        \|
            \bar H_n\bX\Delta_\beta
        \|_1
    \right),
\]
where the first term is $O_P(n\rho_{1,n})$ under Assumption~\ref{assump:pen-scale}-\romannumeral2), and the second term satisfies
\[
    \|
        \bar H_n\bX\Delta_\beta
    \|_1
    \le
    \|
        \bar H_n\bX
    \|_{1,1}
    \|\Delta_\beta\|_1
    = O_P\left(n\cdot \kappa_{1,n}\cdot q\sqrt{\frac{\log p}{n}}\right)
\]
by the last part of Assumption~\ref{assump:graph-curv}-\romannumeral3). It therefore follows that
\begin{equation}
    \frac{1}{n}
    \|\Delta_\eta\|_1
    =
    O_P\left(
        \rho_{1,n}
        +
        \kappa_{1,n}
        q\sqrt{\frac{\log p}{n}}
    \right).
    \label{A.8}
\end{equation}
Again, by triangular inequality,
\[
    \frac{1}{n}
    \|\Delta_\alpha\|_1
    \le
    \frac{1}{n}
    \|\Delta_\eta\|_1
    +
    \frac{1}{n}
    \|\bX\Delta_\beta\|_1,
\]
and since \(\max_{i,j}|X_{ij}|\le R\) by Assumption~\ref{assump:design},
\(
    |X_i^\top\Delta_\beta|
    \le
    R\|\Delta_\beta\|_1.
\)
Consequently,
\begin{equation}
    \frac{1}{n}
    \|\bX\Delta_\beta\|_1
    =
    \frac{1}{n}
    \sum_{i=1}^n
    |X_i^\top\Delta_\beta|
    \le
    R\|\Delta_\beta\|_1
    =
    O_P\left(
        q\sqrt{\frac{\log p}{n}}
    \right).
    \label{A.9}
\end{equation}
Combining \eqref{A.8} and \eqref{A.9} yields
\[
    \frac{1}{n}
    \|
        \boldsymbol\alpha^*-\boldsymbol\alpha^\dagger
    \|_1
    =
    O_P\left(
        \rho_{1,n}
        +
        (1+\kappa_{1,n})
        q\sqrt{\frac{\log p}{n}}
    \right)
\]
which completes the proof.
\end{proof}

\begin{proof}[Proof of Theorem~\ref{thm:consistency}]
Define the estimation error vectors and the corresponding linear predictors as
\(
    \Delta_\beta:=\hat{\bbeta}-\bbeta^*,
    \Delta_\alpha:=\hat{\balpha}-\balphas,
    \Delta_\eta
    :=
    \hat{\boldsymbol\eta}-\boldsymbol\eta^*
    =
    \Delta_\alpha+\bX\Delta_\beta.
\)
By definition of the penalized PMLE,
\[
    \hat\btheta
    =
    \argmin_{\btheta}
    \left\{
        \mathbb P_n\mathcal L(\btheta)
        +
        \tau_n\|\bbeta\|_1
    \right\},
\]
and the optimality of \(\hat\btheta\) implies
\[
    \mathbb P_n\mathcal L(\hat\btheta)
    +
    \tau_n\|\hat\bbeta\|_1
    \le
    \mathbb P_n\mathcal L(\btheta^*)
    +
    \tau_n\|\bbeta^*\|_1.
\]
Adding and subtracting \(\mathbb P_0\mathcal L\), we obtain
\begin{align}
    \mathbb P_0\mathcal L(\hat\btheta)
    -
    \mathbb P_0\mathcal L(\btheta^*)
    &\le
    -
    \left[
        \{\mathbb P_n-\mathbb P_0\}
        \mathcal L(\hat\btheta)
        -
        \{\mathbb P_n-\mathbb P_0\}
        \mathcal L(\btheta^*)
    \right]
    +
    \tau_n
    \left(
        \|\bbeta^*\|_1-\|\hat\bbeta\|_1
    \right).
    \label{eq:basic-ineq}
\end{align}
The empirical fluctuation only comes from the linear term involving
\(\boldsymbol Y\), and recalling the notation
\(
    Z_i=Y_i-\mu_i^0,
\) and \(
    \mu_i^0:=\mathbb E_0Y_i,
\)
we have
\[
    \{\mathbb P_n-\mathbb P_0\}
    \mathcal L(\hat\btheta)
    -
    \{\mathbb P_n-\mathbb P_0\}
    \mathcal L(\btheta^*)
    =
    -
    \frac1n
    \boldsymbol Z^\top\Delta_\eta.
\]
Substituting into \eqref{eq:basic-ineq} yields
\begin{equation}
    \mathbb P_0\mathcal L(\hat\btheta)
    -
    \mathbb P_0\mathcal L(\btheta^*)
    \le
    \frac1n\boldsymbol Z^\top\Delta_\eta
    +
    \tau_n
    \left(
        \|\bbeta^*\|_1-\|\hat\bbeta\|_1
    \right).
    \label{eq:basic-with-Z}
\end{equation}

We next derive the lower bound for the curvature of the smooth population criterion.
By construction,
\begin{align}
    \mathbb P_0\mathcal L(\btheta^*+\Delta)
    -
    \mathbb P_0\mathcal L(\btheta^*)
    &=
    \frac1n
    \sum_{i=1}^n
    \left[
        |\Omega_i|
        \left\{
            \exp(\eta_i^*+\Delta_{\eta,i})
            -
            \exp(\eta_i^*)
        \right\}
        -
        \mu_i^0 \Delta_{\eta,i}
    \right]
    \notag\\
    &\quad
    +
    \frac{\gamma_n}{n}
    \left[
        R(\balphas+\Delta_\alpha;\mathcal G_n)
        -
        R(\balphas;\mathcal G_n)
    \right];
    \label{eq:pop-smooth-diff-raw}
\end{align}
and further by the optimality of $\btheta^*$ with respect to $\mathbb P_0\mathcal L$, we have
\begin{align}
    0
    &=
    \nabla\mathbb P_0\mathcal L(\btheta^*)^\top\Delta
    \notag\\
    &=
    \frac1n
    \sum_{i=1}^n
    \left[
        |\Omega_i|\exp(\eta_i^*)
        -
        \mu_i^0
    \right]
    \Delta_{\eta,i}
    +
    \frac{\gamma_n}{n}
    \nabla R(\balphas;\mathcal G_n)^\top\Delta_\alpha.
    \label{eq:pop-score-zero-direction}
\end{align}
Subtracting \eqref{eq:pop-score-zero-direction} from
\eqref{eq:pop-smooth-diff-raw} yields
\begin{align}
    \mathbb P_0\mathcal L(\btheta^*+\Delta)
    -
    \mathbb P_0\mathcal L(\btheta^*)
    &=
    \frac1n
    \sum_{i=1}^n
    |\Omega_i|
    \left[
        \exp(\eta_i^*+\Delta_{\eta,i})
        -
        \exp(\eta_i^*)
        -
        \exp(\eta_i^*)\Delta_{\eta,i}
    \right]
    \notag\\
    &\quad
    +
    \frac{\gamma_n}{n}
    \left[
        R(\balphas+\Delta_\alpha;\mathcal G_n)
        -
        R(\balphas;\mathcal G_n)
        -
        \nabla R(\balphas;\mathcal G_n)^\top\Delta_\alpha
    \right],
    \label{eq:pop-smooth-bregman}
\end{align}
where we analyze the two terms separately.
For the first term of \eqref{eq:pop-smooth-bregman} induced from the working Poisson likelihood,
\[
    \frac1n
    \sum_{i=1}^n
    |\Omega_i|
    \left[
        \exp(\eta_i^*+\Delta_{\eta,i})
        -
        \exp(\eta_i^*)
        -
        \exp(\eta_i^*)\Delta_{\eta,i}
    \right]
    =
    \frac{1}{n}
    \Delta_\eta^\top
    \bar W_n
    \Delta_\eta,
\]
where
\[
    \bar W_n
    :=
    \int_0^1
    (1-t)
    \operatorname{diag}
    \left\{
        |\Omega_i|
        \exp(\eta_i^*+  \Delta_{\eta,i})
    \right\}_{i=1}^n
    \dd t.
\]
And for the second term of \eqref{eq:pop-smooth-bregman}, by definition of the generalized Hessian in Assumption~\ref{assump:graph-curv},
\[
    \frac{\gamma_n}{n}
    \left[
        R(\balphas+\Delta_\alpha;\mathcal G_n)
        -
        R(\balphas;\mathcal G_n)
        -
        \nabla R(\balphas;\mathcal G_n)^\top\Delta_\alpha
    \right]
    =
    \frac{1}{n}
    \Delta_\alpha^\top
    \bar H_n
    \Delta_\alpha,
\]
where \(\bar H_n\) is the integrated generalized Hessian of
\(\gamma_nR(\btalpha;\mathcal G_n)\) along the segment from \(\balphas\) to
\(\balphas+\Delta_\alpha\).
Plugging back into \eqref{eq:pop-smooth-bregman} yields
\begin{equation}
    \mathbb P_0\mathcal L(\btheta^*+\Delta)
    -
    \mathbb P_0\mathcal L(\btheta^*)
    =
    \frac{1}{n}
    \Delta_\eta^\top
    \bar W_n
    \Delta_\eta
    +
    \frac{1}{n}
    \Delta_\alpha^\top
    \bar H_n
    \Delta_\alpha.
    \label{eq:pop-smooth-bregman-decomp}
\end{equation}

Under the asymptotic regime specified in Definition~\ref{def:asym}, and under Assumption~\ref{assump:bounded-cont-intensity} applied to the $\eta$-neighborhood specified in Assumption~\ref{assump:graph-curv},
there exists a constant \(c_W>0\), not depending on \(n\), such that $\lambda_{\min}(\bar W_n)\ge c_W$
with probability tending to 1. It then follows for the first term in \eqref{eq:pop-smooth-bregman-decomp} that
\begin{equation}
    \frac{1}{n}
    \Delta_\eta^\top
    \bar W_n
    \Delta_\eta
    \ge
    c_W
    \frac{\|\Delta_\eta\|_2^2}{n}.
    \label{eq:poisson-curvature}
\end{equation}
For the second term in \eqref{eq:pop-smooth-bregman-decomp}, recall that $\Delta_\alpha=\Delta_\eta-\bX\Delta_\beta$, we have
\[
    \Delta_\alpha^\top\bar H_n\Delta_\alpha
    =
    \Delta_\eta^\top\bar H_n\Delta_\eta
    -
    2\Delta_\eta^\top\bar H_n\bX\Delta_\beta
    +
    \Delta_\beta^\top\bX^\top\bar H_n\bX\Delta_\beta.
\]
By the positive semi-definite property of $\bar H_n$ in Assumption~\ref{assump:graph-curv}-\romannumeral3), it holds that $\Delta_\eta^\top\bar H_n\Delta_\eta\ge0$. Furthermore, under the \(\ell_2\) fusion compatibility condition of Assumption~\ref{assump:graph-curv}-\romannumeral2),
\[
    \frac1n
    \Delta_\beta^\top
    \bX^\top
    \bar H_n
    \bX
    \Delta_\beta
    \ge
    \kappa_F\|\Delta_\beta\|_2^2.
\]
Combined with Assumption~\ref{assump:graph-curv}\,\romannumeral1), there exists a constant \(C_H<\infty\), not depending on \(n\), such that
\[
    \left\|
        \frac1n
        \bX^\top
        \bar H_n
        \Delta_\eta
    \right\|_\infty
    \le
    C_Ha_n
\]
with probability tending to 1. By Holder's inequality,
\[
    \frac1n
    \left|
        \Delta_\eta^\top
        \bar H_n
        \bX\Delta_\beta
    \right|
    \le
    \left\|
        \frac1n
        \bX^\top
        \bar H_n
        \Delta_\eta
    \right\|_\infty
    \|\Delta_\beta\|_1
    \le
    C_Ha_n\|\Delta_\beta\|_1.
\]
Consequently, with probability converging to 1,
\begin{equation}
    \mathbb P_0\mathcal L(\btheta^*+\Delta)
    -
    \mathbb P_0\mathcal L(\btheta^*)
    \ge
    c_W
    \frac{\|\Delta_\eta\|_2^2}{n}
    +
    \kappa_F
    \|\Delta_\beta\|_2^2
    -
    2C_H a_n\|\Delta_\beta\|_1 .
    \label{eq:curvature-existing}
\end{equation}

Meanwhile, Assumption~\ref{assump:centered-score} states that there exists a constant \(C_Z<\infty\), not depending on \(n\), such that
\[
    \left|
        \frac1n\boldsymbol Z^\top\Delta_\eta
    \right|
    \le
    C_Z
    \frac{\|\Delta_\eta\|_2}{\sqrt n}
    \sqrt q\,a_n
    +
    C_Zqa_n^2
\]
with probability tending to 1. By Young's inequality,
\[
    C_Z
    \frac{\|\Delta_\eta\|_2}{\sqrt n}
    \sqrt q\,a_n
    \le
    \frac{c_W}{2}
    \frac{\|\Delta_\eta\|_2^2}{n}
    +
    \frac{C_Z^2}{2c_W}qa_n^2.
\]
Let
\[
    C_{ZW}:=
    C_Z+\frac{C_Z^2}{2c_W}.
\]
Therefore,
\begin{equation}
    \left|
        \frac1n\boldsymbol Z^\top\Delta_\eta
    \right|
    \le
    \frac{c_W}{2}
    \frac{\|\Delta_\eta\|_2^2}{n}
    +
    C_{ZW}qa_n^2.
    \label{eq:centered-score-absorbed}
\end{equation}

Combining \eqref{eq:basic-with-Z},
\eqref{eq:curvature-existing}, and
\eqref{eq:centered-score-absorbed}, we obtain
\begin{equation}
    \frac{c_W}{2}
    \frac{\|\Delta_\eta\|_2^2}{n}
    +
    \kappa_F
    \|\Delta_\beta\|_2^2
    \le
    C_{ZW}qa_n^2
    +
    2C_H a_n\|\Delta_\beta\|_1
    +
    \tau_n
    \left(
        \|\bbeta^*\|_1-\|\hat\bbeta\|_1
    \right).
    \label{eq:pre-main-ineq}
\end{equation}

Now let $b_n:=\|\bbeta^*_{S^C}\|_1$.
Since \(\bbeta^0_{S^C}=0\), Lemma~\ref{lem:target-param-fusion} and
\(\bbeta^\dagger=\bbeta^0\) imply
\[
    b_n
    \le
    \|\bbeta^*-\bbeta^0\|_1
    =
    O_P\left(
        q\sqrt{\frac{\log p}{n}}
    \right).
\]
Moreover, we also have
\begin{align}
    \|\bbeta^*\|_1-\|\hat\bbeta\|_1
    &=
    \|\bbeta^*_S\|_1+\|\bbeta^*_{S^C}\|_1
    -
    \|\bbeta^*_S+\Delta_{\beta,S}\|_1
    -
    \|\bbeta^*_{S^C}+\Delta_{\beta,S^C}\|_1
    \notag\\
    &\le
    \|\Delta_{\beta,S}\|_1
    -
    \|\Delta_{\beta,S^C}\|_1
    +
    2b_n.
    \label{eq:l1-decomp}
\end{align}
Substituting \eqref{eq:l1-decomp} into \eqref{eq:pre-main-ineq} yields
\begin{align}
    \frac{c_W}{2}
    \frac{\|\Delta_\eta\|_2^2}{n}
    +
    \kappa_F
    \|\Delta_\beta\|_2^2
    +
    \tau_n\|\Delta_{\beta,S^C}\|_1
    &\le
    C_{ZW}qa_n^2
    +
    2C_H a_n\|\Delta_\beta\|_1
    +
    \tau_n\|\Delta_{\beta,S}\|_1
    +
    2\tau_n b_n.
    \label{eq:main-before-absorb}
\end{align}
Decomposing
\[
    \|\Delta_\beta\|_1
    =
    \|\Delta_{\beta,S}\|_1
    +
    \|\Delta_{\beta,S^C}\|_1,
\]
and choosing the lower multiplicative constant in \(\tau_n\asymp a_n\)
large enough so that
\[
    \tau_n\ge 4C_Ha_n,
\]
we may absorb the term
$2C_Ha_n\|\Delta_{\beta,S^C}\|_1$
into the LHS. Continuing from \eqref{eq:main-before-absorb}, we obtain
\begin{align}
    \frac{c_W}{2}
    \frac{\|\Delta_\eta\|_2^2}{n}
    +
    \kappa_F
    \|\Delta_\beta\|_2^2
    +
    \frac{\tau_n}{2}\|\Delta_{\beta,S^C}\|_1
    &\le
    \left(\tau_n+2C_Ha_n\right)
    \|\Delta_{\beta,S}\|_1
    +
    2\tau_n b_n
    +
    C_{ZW}qa_n^2.
    \label{eq:main-ineq-final}
\end{align}
Dropping the nonnegative curvature terms in
\eqref{eq:main-ineq-final} gives
\[
    \|\Delta_{\beta,S^C}\|_1
    \le
    \left(
        2+\frac{4C_Ha_n}{\tau_n}
    \right)
    \|\Delta_{\beta,S}\|_1
    +
    4b_n
    +
    2C_{ZW}\frac{qa_n^2}{\tau_n}.
\]
Using \(\tau_n\asymp a_n\) and \(b_n=O_P(qa_n)\), we obtain
\begin{equation}
    \|\Delta_{\beta,S^C}\|_1
    \le
    C_{\mathrm{cone}}\|\Delta_{\beta,S}\|_1
    +
    O_P(qa_n)
    \label{eq:beta-cone-bound}
\end{equation}
for a constant \(C_{\mathrm{cone}}<\infty\).
Thus, the estimation error $\Delta\in\mathcal C_n(c_1,c_2)$ 
with probability converging to 1, for sufficiently large constants
\(c_1,c_2\).

Next, dropping the term
\(\frac{\tau_n}{2}\|\Delta_{\beta,S^C}\|_1\) from the LHS of
\eqref{eq:main-ineq-final}, we have
\[
    \frac{c_W}{2}
    \frac{\|\Delta_\eta\|_2^2}{n}
    +
    \kappa_F
    \|\Delta_\beta\|_2^2
    \le
    \left(\tau_n+2C_Ha_n\right)
    \|\Delta_{\beta,S}\|_1
    +
    2\tau_n b_n
    +
    C_{ZW}qa_n^2.
\]
By Cauchy's inequality,
\[
    \|\Delta_{\beta,S}\|_1
    \le
    \sqrt q\|\Delta_\beta\|_2.
\]
Therefore,
\[
    \left(\tau_n+2C_Ha_n\right)
    \|\Delta_{\beta,S}\|_1
    \le
    \left(\tau_n+2C_Ha_n\right)
    \sqrt q\|\Delta_\beta\|_2.
\]
By Young's inequality,
\[
    \left(\tau_n+2C_Ha_n\right)
    \sqrt q\|\Delta_\beta\|_2
    \le
    \frac{\kappa_F}{2}\|\Delta_\beta\|_2^2
    +
    \frac{\left(\tau_n+2C_Ha_n\right)^2}{2\kappa_F}q.
\]
Consequently,
\[
    \frac{c_W}{2}
    \frac{\|\Delta_\eta\|_2^2}{n}
    +
    \frac{\kappa_F}{2}
    \|\Delta_\beta\|_2^2
    \le
    2\tau_n b_n
    +
    C_{ZW}qa_n^2
    +
    \frac{\left(\tau_n+2C_Ha_n\right)^2}{2\kappa_F}q.
\]
Using \(\tau_n\asymp a_n\) and \(b_n=O_P(qa_n)\), we obtain
\[
    \frac{\|\Delta_\eta\|_2^2}{n}
    +
    \|\Delta_\beta\|_2^2
    =
    O_P(qa_n^2).
\]
Hence,
\[
    \frac1{\sqrt n}
    \|\hat{\boldsymbol\eta}-\boldsymbol\eta^*\|_2
    =
    O_P(\sqrt q\,a_n),
\]
and
\[
    \|\hat\bbeta-\bbeta^*\|_2
    =
    O_P(\sqrt q\,a_n).
\]

Finally, by \eqref{eq:beta-cone-bound},
\[
    \|\Delta_\beta\|_1
    =
    \|\Delta_{\beta,S}\|_1
    +
    \|\Delta_{\beta,S^C}\|_1
    \le
    \left(1+C_{\mathrm{cone}}\right)\|\Delta_{\beta,S}\|_1
    +
    O_P(qa_n),
\]
where, since
\[
    \|\Delta_{\beta,S}\|_1
    \le
    \sqrt q\|\Delta_\beta\|_2
    =
    O_P(qa_n),
\]
we obtain
\[
    \|\hat\bbeta-\bbeta^*\|_1
    =
    O_P(qa_n).
\]
Also, by Cauchy's inequality,
\[
    \frac1n
    \|\hat{\boldsymbol\eta}-\boldsymbol\eta^*\|_1
    \le
    \frac1{\sqrt n}
    \|\hat{\boldsymbol\eta}-\boldsymbol\eta^*\|_2
    =
    O_P(\sqrt q\,a_n).
\]
\end{proof}

\subsection{Inference}
\label{app:inference}

We first present the additional assumptions required to establish asymptotic normality of $\hat \bb$ explicitly:

\begin{assumption}[Additional regularity conditions for inference]
\label{assump:inference}
Let \(\mathcal J\subseteq\{1,\ldots,p\}\) denote the coordinates for which
inference is conducted. For any \(\boldsymbol\eta\in\mathbb R^n\), define
\[
    W(\boldsymbol\eta)
    :=
    \operatorname{diag}
    \left\{
        |\Omega_i|\exp(\eta_i)
    \right\}_{i=1}^n,
    \qquad
    H(\boldsymbol\eta)
    :=
    \frac1n\bX^\top W(\boldsymbol\eta)\bX.
\]
Let $\eta_i^*:=\alpha_i^*+X_i^\top\bbeta^*$, $ \mu_i^*:=|\Omega_i|\exp(\eta_i^*)$ and $ Z_i^*:=Y_i-\mu_i^*$,
with vectorized form $\boldsymbol\eta^*:=(\eta_1^*,\ldots,\eta_n^*)^\top$ and $\boldsymbol Z^*:=(Z_1^*,\ldots,Z_n^*)^\top$.
Define $\Omega_n:=\frac1n\operatorname{Var}_0(\bX^\top\boldsymbol Z^*)$ where $\text{Var}_0(\cdot)$ indicates variance with respect to the true data generating mechanism. We further introduce the short-hand notations $W^*:=W(\boldsymbol\eta^*)$ and $ H^*:=H(\boldsymbol\eta^*)$.

For each \(j\in\mathcal J\), let \(m_j\in\mathbb R^p\) be the solution to the population-level version of \eqref{equ:m-debias}. We assume the following conditions to hold:

\begin{itemize}
    \item[\romannumeral1)]
    \textbf{Strengthened empirical-process bound.}
We require $a_n$, defined in Assumption~\ref{assump:centered-score}, to satisfy $\sqrt n\,q a_n^2=o(1)$. 
Moreover, there exists a deterministic sequence \(\zeta_n\to0\) such that
\[
    \max_{j\in\mathcal J}
    \|H^*m_j-e_j\|_\infty
    \le
    \zeta_n,
    \text{ and }
    \sqrt n\,\zeta_n q a_n=o(1).
\]
    \item[\romannumeral2)]
    \textbf{Local Hessian stability.}
    Let $r_n$ be a sequence such that $r_n\asymp\sqrt q\,b_n$ where $b_n:=a_n+\sqrt{\frac{\log p}{n}}$. Define
    \[
        \mathcal E_n(r_n)
        :=
        \left\{
            \boldsymbol\eta\in\mathbb R^n:
            \frac1{\sqrt n}
            \|\boldsymbol\eta-\boldsymbol\eta^*\|_2
            \le r_n
        \right\},
    \]
    then we require there exists a sequence \(\delta_{H,n}\to0\) such that $\sqrt n\,\delta_{H,n}q a_n=o(1)$ and
    \[
        \max_{j\in\mathcal J}
        \sup_{\boldsymbol\eta\in\mathcal E_n(r_n)}
        \|
            \{H(\boldsymbol\eta)-H^*\}m_j
        \|_\infty
        \le
        \delta_{H,n}.
    \]

    \item[\romannumeral3)]
    \textbf{Nuisance-intercept condition.}
    Let \(\mathcal A_n(s_n)\subseteq\mathbb R^n\) denote a deterministic local class of intercept perturbations around zero. 
    We assume
    \[
        \max_{j\in\mathcal J}
        \sup_{\substack{
            \boldsymbol a\in\mathcal A_n(s_n)\\
            \boldsymbol\eta\in\mathcal E_n(r_n)
        }}
        \left|
            \frac1{\sqrt n}
            m_j^\top
            \bX^\top
            W(\boldsymbol\eta)
            \boldsymbol a
        \right|
        =
        o(1).
    \]

    \item[\romannumeral4)]
    \textbf{Strengthened graph fusion.}
    Recall $G_n^\dagger
        :=
        \gamma_n\nabla_\balpha R(\balphad;\mathcal G_n)$
    from Assumption~\ref{assump:pen-scale}, and let $K_n^\dagger
        :=
        \frac1n
        \bX^\top
        \bar H_n^\dagger
        \bX$,
    where \(\bar H_n^\dagger\) denotes the local generalized Hessian of
    \(\gamma_nR(\cdot;\mathcal G_n)\) in a neighborhood of \(\balphad\). We assume the following holds uniformly in this neighborhood of \(\balphad\):
    \[
        \max_{j\in\mathcal J}
        \frac{1}{\sqrt n}
        \left|
            e_j^\top
            (K_n^\dagger)^{-1}
                \bX^\top G_n^\dagger
        \right|
        =
        o(1).
    \]

    \item[\romannumeral5)]
    \textbf{Variance regularity.}
    For each \(j\in\mathcal J\), define
    \[
        \sigma_j^2
        :=
        m_j^\top\Omega_n m_j
        =
        \frac1n
        \operatorname{Var}_0
        \left(
            m_j^\top\bX^\top\boldsymbol Z^*
        \right)
        =
        \frac1n
        \sum_{i,k=1}^n
        (m_j^\top X_i)(m_j^\top X_k)
        \operatorname{Cov}_0(Z_i^*,Z_k^*).
    \]
    We assume $0<c_\sigma
        \le
        \min_{j\in\mathcal J}\sigma_j^2
        \le
        \max_{j\in\mathcal J}\sigma_j^2
        \le
        C_\sigma<\infty$. 
\end{itemize}
\end{assumption}

\begin{remark}
Conditions \romannumeral1) and \romannumeral2) establish strengthened error rates that guarantee $o_P(n^{-1/2})$ convergence on top of what
Theorem~\ref{thm:consistency} established.
Condition \romannumeral4) strengthens Assumption~\ref{assump:pen-scale}, which only controls \(\|n^{-1}\bX^\top G_n^\dagger\|_\infty\) at
the high-dimensional estimation-noise scale, so that the gap between $\bbeta^*$ and the target parameter $\bbeta^0$ is controlled at 
the \(n^{-1/2}\) scale. 
A sufficient condition for \romannumeral4) is
    \[
        \frac1n\left\|
            \bX^\top G_n^\dagger
        \right\|_\infty
        =
        o(n^{-1/2}),
    \]
    together with a bounded-inverse condition for \(K_n^\dagger\). For
    \(\ell_2\) fusion, where \(G_n^\dagger=\gamma_n\tilde L_n\balphad\), this
    reduces to
    \[
        \frac{\gamma_n}{n}\left\|
            \bX^\top\tilde L_n\balphad
        \right\|_\infty
        =
        o(n^{-1/2}).
    \]
\end{remark}

\begin{proof}[Proof of Theorem~\ref{thm:asym-normal}]
    Recall that the de-biased estimator is defined as
    \[
        \hat \bb = \hat\bbeta + \frac{1}{n}M X^\top\left[ \bY - \bB\odot \exp\left( \hat\balpha + \bX\hat\bbeta \right)\right].
    \]
For each $j$, by definition of \(\hat b_j\),
\[
    \hat b_j-\beta_j^*
    =
    \hat\beta_j-\beta_j^*
    +
    \frac1n
    m_j^\top\bX^\top
    \left(
        \bY-\hat{\boldsymbol\mu}
    \right).
\]
Adding and subtracting
\(\boldsymbol\mu^*=(\mu_1^*,\ldots,\mu_n^*)^\top\), we obtain
\[
    \hat b_j-\beta_j^*
    =
    \frac1n
    m_j^\top\bX^\top\boldsymbol Z^*
    +
    (\hat\beta_j-\beta_j^*)
    -
    \frac1n
    m_j^\top\bX^\top
    (\hat{\boldsymbol\mu}-\boldsymbol\mu^*).
\]
By the mean-value theorem applied element-wise to
\(\mu_i(\eta_i)=|\Omega_i|\exp(\eta_i)\), we have 
$ \hat{\boldsymbol\mu}-\boldsymbol\mu^*
    =
    \bar W
    (\hat{\boldsymbol\eta}-\boldsymbol\eta^*)$,
where
\[
    \bar W
    =
    \operatorname{diag}
    \left\{
        |\Omega_i|
        \int_0^1
        \exp[
            \eta_i^*
            +
            t(\hat\eta_i-\eta_i^*)
        ]
        \dd t
    \right\}_{i=1}^n.
\]
Therefore,
\[
    \hat b_j-\beta_j^*
    =
    \frac1n
    m_j^\top\bX^\top\boldsymbol Z^*
    +
    (\hat\beta_j-\beta_j^*)
    -
    \frac1n
    m_j^\top\bX^\top
    \bar W
    \left[
        (\hat\balpha-\balphas)
        +
        \bX(\hat\bbeta-\bbeta^*)
    \right].
\]
Rearranging yields
\[
    \hat b_j-\beta_j^*
    =
    \frac1n
    m_j^\top\bX^\top\boldsymbol Z^*
    +
    R_{1j}
    +
    R_{2j},
\]
where we analyze the remainder terms
\[
    R_{1j}
    :=
    e_j^\top(\hat\bbeta-\bbeta^*)
    -
    m_j^\top
    \left(
        \frac1n\bX^\top\bar W\bX
    \right)
    (\hat\bbeta-\bbeta^*),
\]
and
\[
    R_{2j}
    :=
    -
    \frac1n
    m_j^\top\bX^\top
    \bar W
    (\hat\balpha-\balphas)
\]
separately. For \(R_{1j}\), define $\bar H
    :=
    \frac1n\bX^\top\bar W\bX$, then
\[
    |R_{1j}|
    \le
    \|e_j-\bar Hm_j\|_\infty
    \|\hat\bbeta-\bbeta^*\|_1.
\]
By Assumption~\ref{assump:inference}\romannumeral1) and \romannumeral2),
\[
    \|e_j-\bar Hm_j\|_\infty
    \le
    \|e_j-H^*m_j\|_\infty
    +
    \|(\bar H-H^*)m_j\|_\infty
    \le
    \zeta_n+\delta_{H,n},
\]
when \(\hat{\boldsymbol\eta}\in\mathcal E_n(r_n)\), which holds true with
probability tending to 1 by Theorem~\ref{thm:consistency}. By Theorem~\ref{thm:consistency} and Assumption~\ref{assump:inference}\romannumeral4), it holds that
\[
    \|\hat\bbeta-\bbeta^0\|_1=O_P(qb_n),
    \qquad
    \|\hat\bbeta-\bbeta^0\|_2=O_P(\sqrt q\,b_n).
\]
Moreover, since
\[
    \|\hat\bbeta-\bbeta^*\|_1
    \le
    \|\hat\bbeta-\bbeta^0\|_1
    +
    \|\bbeta^*-\bbeta^0\|_1,
\]
we also have
\[
    \|\hat\bbeta-\bbeta^*\|_1=O_P(qb_n).
\]
Therefore,
\[
    \sqrt n\,|R_{1j}|
    \le
    \sqrt n\,
    \|e_j-\bar Hm_j\|_\infty
    \|\hat\bbeta-\bbeta^*\|_1
    =
    O_P\left[
        \sqrt n(\zeta_n+\delta_{H,n})qb_n
    \right]
    =
    o_P(1),
\]
by Assumption~\ref{assump:inference}\romannumeral1) and \romannumeral2).
Further, by Assumption~\ref{assump:inference}-\romannumeral3),
\[
    \sqrt n\,|R_{2j}|
    =
    \left|
        \frac1{\sqrt n}
        m_j^\top\bX^\top
        \bar W
        (\hat\balpha-\balphas)
    \right|
    =
    o_P(1),
\]
again since the intermediate linear predictor in \(\bar W\) falls in
\(\mathcal E_n(r_n)\) with probability tending to 1.
Hence
\[
    \sqrt n(\hat b_j-\beta_j^*)
    =
    \frac1{\sqrt n}
    m_j^\top\bX^\top\boldsymbol Z^*
    +
    o_P(1),
\]
and by Assumption~\ref{assump:inference}\romannumeral5), yields
\[
    \frac{
        \sqrt n(\hat b_j-\beta_j^*)
    }{
        \sigma_j
    }
    \xrightarrow{\mathrm d}
    N(0,1).
\]
Finally, due to Assumption~\ref{assump:inference}-\romannumeral4), it also holds that
\[
    \sqrt n|\beta_j^*-\beta_j^0|=o(1).
\]
Applying Slutsky's theorem leads to
\[
    \frac{
        \sqrt n(\hat b_j-\beta_j^0)
    }{
        \sigma_j
    }
    \xrightarrow{\mathrm d}
    N(0,1)
\]
which establishes Theorem~\ref{thm:asym-normal}.
    
To see why the covariance estimator $\hat\bSigma$ in (\ref{equ:Sig-hat}) is a conservative estimate for $\mathbb E_0\nabla_\bbeta \ell(\balpha^\dagger, \bbeta^0)\nabla_\bbeta \ell(\balpha^\dagger, \bbeta^0)^\top$, note that
     \begin{align}
    \notag \frac{1}{n}\sum_{i=1}^n X_i^\top X_i & {\text{Var}}(Y_i\mid X_i) 
     = \frac{1}{n}\sum_{i=1}^n X_i^\top X_i \mathbb E_{\varepsilon_i^*} \mathbb E_{Y_i\mid\varepsilon_i^*} (Y_i - \lvert\Omega_i\rvert P_i \exp(\tilde\alpha_i + X_i\bbeta^0 + \varepsilon_i^*))^2 \\
    \notag & \preceq \frac{2}{n}\sum_{i=1}^n X_i^\top X_i\left[ \left(Y_i - \lvert\Omega_i\rvert P_i \exp(\hat\alpha_i + X_i\hat\bbeta)\right)^2 + 
    \mathbb E_{\varepsilon_i^*}\left(\lvert\Omega_i\rvert P_i \exp(\hat\alpha_i + X_i\hat\bbeta) - \lvert\Omega_i\rvert P_i \exp(\tilde\alpha_i + X_i\bbeta^0 + \varepsilon_i^*)\right)^2 \right]
    \end{align} 
    where $\varepsilon_i^* = \varepsilon(s_i^*)$ for the location $s_i^*$ defined in Lemma~\ref{lem:target-param}, and we recall that $\alpha_i^\dagger = \tilde\alpha_i + (X(s_i^*)-X_i)\bbeta + \log\mathbb E_0[\exp(\varepsilon_i^*)]$.
    

\end{proof}

\end{document}